\documentclass[11pt]{article}
\pdfoutput=1

\usepackage{fullpage}
\usepackage{amsmath}
\usepackage{amssymb}
\usepackage{setspace}
\usepackage{bbm}
\usepackage{dsfont}
\usepackage{graphics}
\usepackage{cite}
\usepackage{hyperref}

\hypersetup{
    bookmarks=true,%
    colorlinks,%
    citecolor=blue,%
    filecolor=blue,%
    linkcolor=blue,%
    urlcolor=blue
}

\usepackage[font=footnotesize,labelfont=bf,justification=centerlast,width=.94\textwidth]{caption}

\usepackage{color}

\usepackage[normalem]{ulem}
\usepackage{epsfig}

\newcommand{\be}{\begin{equation}}
\newcommand{\ee}{\end{equation}}
\newcommand{\bes}{\begin{equation*}}
\newcommand{\ees}{\end{equation*}}
\newcommand{\bea}{\begin{eqnarray}}
\newcommand{\eea}{\end{eqnarray}}
\newcommand{\beas}{\begin{eqnarray*}}
\newcommand{\eeas}{\end{eqnarray*}}

\newcommand{\p}{\partial}

\newcommand{\bmat}{\begin{bmatrix}}
\newcommand{\emat}{\end{bmatrix}}

\newcommand{\slt}{SL(2,\RR)}

\newcommand{\RR}{\mathbb{R}}

\def\tr{{\rm Tr}}
\def\Tr{{\rm Tr}}
\def\le{\left}
\def\ri{\right}
\def\ep{{\epsilon}}
\newcommand\sO{{\ensuremath{{\mathcal O}}}}

\def\Tr{{\rm Tr}}
\def\le{\left}
\def\ri{\right}

\def\le{\left}
\def\ri{\right}
\def\ha{{1\over 2}}

\def\al{{\alpha}}

\def\tr{{\rm Tr}}
\def\Tr{{\rm Tr}}

\def\th{{\theta}}

\def\ep{{\epsilon}}

\newcommand\ga{{\ensuremath{{\gamma}}}}

\newcommand\Sig{\Sigma}
\newcommand\lam{\lambda}

\def\lam{{\lambda}}

\def\eeq{\end{equation}}

\def\Tr{\mathop{\rm Tr}}

\newcommand\sD{{\ensuremath{{\mathcal D}}}}

\newcommand\sH{{\ensuremath{{\mathcal H}}}}

\newcommand\sR{{\ensuremath{{\mathcal R}}}}

\newcommand\bL{{\overline L}}
\newcommand\bl{{\overline \ell}}
\newcommand\ba{{\overline a}}
\newcommand\btau{{\overline \tau}}
\newcommand\bth{{\overline \theta}}

\def\th{{\theta}}

\newcommand\bB{{\overline{B}}}

\begin{document}
\numberwithin{equation}{section}
{
\begin{titlepage}
\begin{center}

\hfill \\
\hfill \\
\vskip 0.6in

{\Large \bf Entanglement Entropy in Warped Conformal Field Theories}\\

\vskip 0.4in

{\large  Alejandra Castro, Diego M. Hofman, and Nabil Iqbal}\\

\vskip 0.3in

{\it Institute for Theoretical Physics, University of Amsterdam,
Science Park 904, Postbus 94485, 1090 GL Amsterdam, The Netherlands} \vskip .5mm

\end{center}

\vskip 0.45in

\begin{center} {\bf ABSTRACT } \end{center}

We present a detailed discussion of entanglement entropy in $(1+1)$-dimensional Warped Conformal Field Theories (WCFTs). We implement the Rindler method to evaluate entanglement and Renyi entropies for a single interval and along the way we interpret our results in terms of twist field correlation functions. Holographically a WCFT can be described in terms of Lower Spin Gravity, a $SL(2, \mathbb{R}) \times U(1)$ Chern-Simons theory in three dimensions. We show how to obtain the universal field theory results for entanglement in a WCFT via holography. For the geometrical description of the theory we introduce the concept of geodesic and massive point particles in the warped geometry associated to Lower Spin Gravity. In the Chern-Simons description we evaluate the appropriate Wilson line that captures the dynamics of a massive particle. 

%

\end{titlepage}
}

\newpage

\tableofcontents
\newpage 
\section{Introduction}
Entanglement entropy is a very useful tool for organizing our understanding of the correlation structure of quantum mechanical systems. In addition to being interesting on purely field-theoretical grounds, one of its recent applications is to the study of holographic duality via the Ryu-Takayanagi formula \cite{Ryu:2006bv,Ryu:2006ef}. This states that in a quantum field theory with a gravity dual, the entanglement entropy of a spatial subregion can be related to a simple geometric object in the bulk, e.g. a minimal area in the simplest case of Einstein gravity. This remarkable prescription relates two very primitive objects -- quantum entanglement and geometry -- on the two sides of the duality, suggesting that a refined understanding of the emergence of a holographic spacetime may eventually be through the entanglement properties of its field-theoretical dual \cite{VanRaamsdonk:2010pw,Swingle:2009bg,Maldacena:2013xja}.   

However, entanglement entropy is also a notoriously difficult quantity to calculate in field theory alone, and there are few exact results available for entanglement entropy in quantum field theory. One of these known results is that for the entanglement entropy of a single interval in the vacuum of a two-dimensional conformal field theory, where there is a celebrated universal formula that applies to the vacuum on the cylinder of {\it any} two dimensional Conformal Field Theory (CFT) \cite{Holzhey:1994we,Calabrese:2004eu,Calabrese:2009qy}:
\be
S_{\rm EE} = \frac{c}{3} \log \left( \frac{L}{\pi \epsilon} \sin \frac{\pi \ell}{L}\right),  \label{eeCFT2}
\ee
with $\ell$ the length of the interval, $L$ the length of the circle on which the theory is defined, and $\ep$ a UV cutoff. 
This formula exists because the global $\slt \times \slt$ invariance of the conformal vacuum in a CFT$_2$ is enhanced to two copies of an infinite-dimensional Virasoso algebra, greatly constraining the dynamics and permitting the existence of universal formulas such as \eqref{eeCFT2} (and, not unrelatedly, the similarly universal Cardy formula for the thermodynamic entropy in a high-temperature state). 

More recently, however, there has been a great deal of study of a {\it different} class of similarly constrained field theories, called Warped Conformal Field Theories (WCFTs) \cite{Hofman:2011zj,Detournay:2012pc}. These WCFTs possess a vacuum that is invariant only under a global $\slt \times U(1)$, which is then enhanced to a single Virasoro and a Kac-Moody algebra. Though these theories are non-relativistic, they possess a similarly infinite-dimensional symmetry group as a standard two-dimensional CFT and indeed there exist notions of modular invariance that permit the derivation of a Cardy-type formula for the the high-energy density of states \cite{Detournay:2012pc}. This is quite remarkable given that these types of results are scarce for non-relativistic theories (see however \cite{Bagchi:2014iea, Hosseini:2015gua,Hosseini:2015uba} for recent work on the structure of entanglement entropy in other non-relativistic field theories). As such, WCFTs offer a range of applications in condensed matter systems, particularly for Quantum Hall states \cite{Ryu:2012he,Son:2013rqa}. Particular examples of WCFTs were constructed in \cite{Hofman:2014loa}, including the very simple theory of a complex free (massive) Weyl fermion.

Given the infinite dimensional symmetry algebra enjoyed by WCFTs, one might then expect the existence of a universal formula for the entanglement entropy of an interval in the vacuum, similarly to the well studied CFT case. In this paper we study this question in detail. 

One of our main results is the derivation, using WCFT techniques, of a formula analogous to \eqref{eeCFT2}, which we present here:
\be
S_{\rm EE} = i P_0^{\rm vac} \ell \left( \frac{\bar L}{L} - \frac{ \bar \ell}{\ell}\right) -4 L_0^{\rm vac} \log \left( \frac{L}{\pi \epsilon} \sin \frac{\pi \ell}{L}\right)~. \label{eewcft} 
\ee
where $\ell$ and ${\bar \ell}$ are the separation of the endpoints of the interval in question in ``space'' and in ``time'' respectively, and $L$ and ${\bar L}$ are related to the identification pattern of the circle that defines the vacuum of the theory. The non-relativistic nature of the theory is evident in this formula, and the precise meaning of space and time in this context will be made clear later. We note also that the answer is naturally expressed in terms of the charges of the vacuum of the theory, and not in terms of the central charge. The same is actually secretly true of the canonical result \eqref{eeCFT2}, as we will explain. 

We turn then to a holographic description of warped CFTs and describe the appropriate generalization of the Ryu-Takayanagi formula. While there are bulk solutions \cite{Anninos:2008fx,Anninos:2008qb,Guica:2011ia,Compere:2013bya} to Einstein gravity (supplemented with other fields or a gravitational Chern-Simons term in the action), that geometrically have a piece that is warped AdS$_3$ and so should be dual to a warped CFT, these theories also possess a great deal of additional and unnecessary structure. The minimal holographic description -- which should be understood as being related to WCFT in the same way that Einstein-AdS$_3$-gravity with no extra fields is related to CFT$_2$ -- has been more recently understood in \cite{Hofman:2014loa}. This involves some novel geometric ideas (that we review below), and can appropriately be called Lower Spin Gravity, as it involves the geometrization of $\slt \times U(1)$ rather than two copies of $\slt$. 

Note now that in three bulk (and two boundary) dimensions the Ryu-Takayanagi formula relates entanglement to the length of a bulk geodesic, which is equivalent to the action of a massive particle moving in the bulk. To understand its analog for holographic warped CFTs we will then need to understand how to couple massive particles to a background metric in Lower Spin Gravity and study the resulting geodesic motion. We perform this first in a metric formulation of Lower Spin Gravity, where we construct the worldline action of a massive particle moving in the bulk. We also describe the computation of entanglement entropy in a Chern-Simons formulation of the theory. This requires a generalization of the Wilson line prescription developed in \cite{Ammon:2013hba} for AdS$_3$ gravity, where the representation space required for the Wilson line is now generated by an auxiliary quantum mechanical system living on the coset $\slt / SO(1,1)$. In both cases we reproduce the field-theoretical results quoted above from a holographic analysis. 

We now present a brief summary of the paper. In Section \ref{app:wcft} we describe the symmetry structure of warped conformal field theories. Much of this material has appeared in the literature before, but Section \ref{subsec:anom} presents a novel covariant description of the Virasoro-Kac-Moody algebra which is helpful for understanding the preferred coordinate axes that are part of the definition of a WCFT. In Section \ref{sec:eeqft} we use warped conformal mappings to derive universal formulas such as \eqref{eewcft} for the entanglement entropy in the vacuum (and states related to it by conformal transformations, such as the finite temperature state). We also interpret our results in terms of twist fields, deriving expressions for their conformal dimensions and $U(1)$ charges.  In Section \ref{sec:holgeom} we explain how to couple massive particles to Lower Spin Gravity and re-interpret the resulting geometric structures as entanglement entropy. In Section \ref{sec:holCS} we study the same problem in the Chern-Simons description of Lower Spin Gravity, where we evaluate the appropriate Wilson line. We conclude with a discussion and some directions for future research in Section \ref{sec:conc}.


\section{Basic properties of WCFT}\label{app:wcft}

We start by gathering some basic properties of Warped Conformal Field Theories. The following equations are based on the results in  \cite{Hofman:2011zj,Detournay:2012pc,Castro:2015uaa}; the reader familiar with these results can skip portions of this section. We will also review and extend some results in \cite{Hofman:2014loa} in section \ref{subsec:anom}: the emphasis here is to explain and  highlight some geometrical properties of WCFTs. 

Consider a (1+1) dimensional theory defined on a  plane which we describe in terms of two coordinates $(z,w)$. On this plane, we denote as $T(z)$ the operator that generates infinitesimal coordinate transformations in $z$ and $P(z)$ the operator that generates $z$ dependent infinitesimal translations in $w$. We can think of these transformations as finite coordinate transformations
\be\label{app1}
z ~\to ~ z=f(z')   ~,\quad w~ \to~ w=w' + g(z')~.
\ee
Classical systems which are invariant under these transformation are known as Warped Conformal Field Theories (WCFTs).

At the quantum level, we define (in CFT language) $T(z)$ as the right moving energy momentum tensor and $P(z)$ as a right moving $U(1)$ Kac-Moody current.  We can define charges
\be
L_n =-{i\over 2\pi}\int dz\, \zeta_n(z) T(z)~,\quad P_n =-{1\over 2\pi}\int dz\, \chi_n(z) P(z)~,
\ee
where we choose the test functions as $\zeta_n=z^{n+1}$ and $\chi_n=z^n$. In terms of the plane charges $(L_n,P_n)$ the commutation relations are
\begin{eqnarray}
\label{eq:canonicalgebra}
[ L_n, L_{n'}] &=&(n-n') L_{n+n'}+\frac{c}{12}n(n^2-1)\delta_{n,-n'}~, \nonumber \\
\, [ L_n, P_{n'}] &=&-n'  P_{n'+n}~, \nonumber \\
\, [ P_n, P_{n'}] &=&k \frac{n}{2}\delta_{n,-n'}~,
\end{eqnarray}
which is a Virasoro-Kac-Moody algebra with central charge $c$ and level $k$. The finite transformation properties of the currents are
 \bea\label{app:finite}
P'(z') &=& {\partial z\over \partial z'}\le(P(z)+{k\over 2}{\partial w'\over \partial z} \ri)~,\cr
T'(z') &=& \le({\partial z\over \partial z'}\ri)^2\le(T(z)-{c\over 12}\{z',z\} \ri) + {\partial z\over \partial z'}{\partial w\over \partial z'} P(z) - {k\over 4}\le({\partial w\over \partial z'}\ri)^2~,
\eea
where
\be
\{z',z \}= {{\partial^3 z'\over \partial z^3}\over {\partial z'\over \partial z}} -{3\over 2}\le({{\partial^2 z'\over \partial z^2}\over {\partial z'\over \partial z}}\ri)^2~.
\ee

Among these finite transformations, there is one that is rather interesting. Consider doing a tilt of the $w$ direction:
\be
z=z'~,\quad w= w' + 2\gamma z'~.
\ee
Under this tilt, the currents transform as
\bea
P'(z') &=& P(z)-{k}\gamma~,\cr
T'(z') &=& T(z) -2\gamma P(z) - {k}\gamma^2~,
\eea
which implies that the modes on the plane transform as
\bea
\label{eq:spectralflow}
 L_n &\to & L_n^{(\gamma)}= L_n+2\gamma\,  P_n+{\gamma^2} \,k\, \delta_{n,0}~,\cr
 P_n &\to & P_n^{(\gamma)}=P_n +\gamma \,k\, \delta_{n,0}~.
\eea
This is the usual spectral flow transformation, which leaves the commutation relations \eqref{eq:canonicalgebra} invariant.

For most of our manipulations, we will be interested in computing observables on the real time cylinder. Given that $(z,w)$ defined the coordinates on the plane, the transformation that takes us back to the cylinder is 
\be\label{zcoor}
z= e^{-i x} ~, \quad w =  t + 2\alpha x~,
\ee
where on the cylinder, $x$ is the  $SL(2,\mathbb{R})$ scaling coordinate and $t$ is the $U(1)$ axis; $\alpha$ is a constant tilt that controls how we define a space quantization slice in our cylinder relative to operator insertions on the plane. The modes of the cylinder are related to the modes on the planes as 
\be
P^{\rm cyl}_n = P_n + \alpha\, k \delta_{n,0}~,\quad L^{\rm cyl}_n = L_n + 2\alpha\, P_n +\le( \alpha^2\, k -{c\over 24}\ri)\delta_{n,0}~.
\ee

%
%


\subsection{Modular Properties of WCFT}

Modular properties of partition functions (and density matrices) will be important in following sections when evaluating entanglement entropy. Here we review the transformation properties of such functions. 

Consider placing a WCFT on a torus. One way to proceed is to have $x \sim x +2\pi$, which defines a particular spatial cycle, and to introduce temperature and angular potential by identifying imaginary time appropriately, which defines the temporal cycle. However, in a WCFT, this choice is not completely equivalent to choosing other spatial cycles. Therefore, the way to proceed is to define a more general torus defined by the following identification 
\be
(x, t) \sim (x - 2\pi a, t + 2\pi \bar a) \sim (x - 2\pi \tau, t + 2 \pi \bar{\tau})~,
\ee
where we introduce $(\bar a, a)$ to allow for any choice of spatial cycle, and $(\bar\tau, \tau)$ are the thermodynamic potentials for  $(P_0^{\rm cyl}, L_0^{\rm cyl})$. The reality properties of $(\tau,\bar\tau)$ depend on how we Wick rotate back to real time. 
For this parametrization of the torus, the partition function reads
\be\label{eq:z1}
Z_{\bar a| a} (\bar \tau| \tau)= {\rm Tr}_{\bar a| a} \le(e^{2 \pi i \bar \tau P_0^{\rm cyl}} e^{- 2\pi i \tau L_0^{\rm cyl}}\ri)~.
\ee
With this notation it is rather simple to relate partition functions labelled by different choices of $(\bar a, a)$. In particular, if we do the change of coordinates
\be\label{eq:uv}
\hat u= {x\over a}~,\quad \hat v=t+{\bar a \over a}x~,
\ee
the relation between the partition functions using the $(x,t)$ coordinates and $(\hat u,\hat v)$ is
\be\label{spectral}
Z_{\bar a |  a} (\bar \tau| \tau) = e^{\pi i k \bar a \left( \bar \tau -  \frac{\tau \bar a}{2 a} \right)} Z_{0|1} (\bar \tau - \frac{\bar a \tau}{a}| \frac{\tau}{a})~.
\ee
Note that we have kept track of the appropriate anomalies, since the coordinate transformation \eqref{eq:uv} that relates  $Z_{\bar a |  a}$ and $Z_{0 | 1 }$ is a spectral flow transformation.  As shown in \cite{Castro:2015uaa}, the modular properties of  partition functions are sensitive to the torus parametrization --the system is not Lorentz invariant after all. From this stand point, we denote the partition function with $(\bar a, a)=(0,1)$ as {\it canonical} as it is calculated on the {\it canonical circle}. We define
\be\label{eq:defhat}
\hat Z(z|\tau) \equiv Z_{0|1}(\bar \tau - \frac{\bar a \tau}{a}|\tau)~,\quad z=\bar \tau - \frac{\bar a \tau}{a}~.
\ee
The function $\hat Z$ has {\it canonical} modular transformation properties. More concretely,  $S$ is the modular transformation that exchanges the spatial and thermal cycles, and invariance of the partition function under $S$ is equivalent to 
\be
 Z_{0|1} (\bar \tau|\tau) = Z_{\bar \tau| \tau} (0|-1)~.
\ee
 This condition  implies that 
\be\label{eq:shz}
 \hat Z (z |\tau) =e^{\pi i k \frac{z^2}{2 \tau}} \hat Z (\frac{z}{\tau} | -\frac{1}{\tau})~.
\ee

Notice that partition functions defined for other choices $(\bar a, a)$ will not satisfy this simple rule. For a complete treatment of these transformations and the restrictions they impose on the theory  see \cite{Castro:2015uaa}.

\subsection{Quantum anomalies and preferred axes in WCFT}\label{subsec:anom}

This section is based on (and extends) results in \cite{Hofman:2014loa}. Our goal is to explain in detail why WCFTs have two preferred axes in space-time;  this will allow us to pick preferred coordinates in space-time given by these axes which justifies our parametrizations of the system in later sections.

As it was stressed before, WCFTs are non relativistic quantum field theories. As such they do not naturally couple to background Riemannian geometry. In \cite{Hofman:2014loa} it was explained that the natural geometric structure in this case corresponds to a form of warped geometry which is, in two dimensions, a type of Newton-Cartan geometry (see for example  \cite{Andringa:2010it,Christensen:2013lma,Christensen:2013rfa,Hartong:2014oma,Hartong:2014pma,Bergshoeff:2014uea,Geracie:2014nka}).  The main point is that WCFTs posses a natural symmetry associated to generalized boosts (sometimes called Carrollian boosts \cite{Bekaert:2015xua,Hartong:2015xda}):
\be\label{eq:boost}
t \rightarrow t + v x ~.
\ee
This symmetry plays the same role in WCFTs that Lorentz boosts play in CFTs; this can readily be seen from the manipulations around (\ref{zcoor}) and (\ref{eq:uv}). Therefore, WCFTs couple to two dimensional geometries where the local symmetry of space-time is given by translations and the boost symmetry (\ref{eq:boost}). The resulting notion of geometry was described in detail in \cite{Hofman:2014loa}: it turns out that the most efficient way to describe this geometry is in the Cartan formalism, where the symmetries act explicitly in tangent space.

Let us look at tangent space invariant tensors. In two dimensional Lorentz invariant theories the first invariant tensor of the geometry is the flat space metric $\eta_{a b}$ where $a,b = t, x$. In contrast,  for warped geometry there are one index objects that are invariant under the boost symmetry, which are defined as follows. Consider, in a covariant language, the position vector
\be
x^a = \left(\begin{array}{c} x \\ t \end{array}\right)~,
\ee
and the boost transformation 
\be
\Lambda^a_{\phantom{a}b} =  \left( \begin{array}{cc} 1 & 0 \\ v & 1 \end{array}\right) ~.
\ee
There exists an invariant vector ($\bar q^a$) and an invariant one-form ($q_a$) given by
\be
\bar q^a =\left( \begin{array}{c} 0 \\ 1 \end{array} \right) \quad\quad  \textrm{and} \quad \quad q_a = \left( \begin{array}{cc} 1 & 0 \end{array}\right) .\label{invvec}
\ee
From these objects we may also define tensor invariants: a degenerate metric $g_{ab}$ and the antisymmetric tensor $h_{ab}$, which are 
\be
g_{ab} \equiv q_a q_b~, \qquad q_a \equiv h_{ab} \bar{q}^b \ . \label{2dtensors}
\ee
These tensors permit two different notions of inner products between two vectors $U, V$:
\be
U \cdot V \equiv U^a V^b g_{ab}  = \left(U^a q_a\right) \left(V^a q_a\right)~,\qquad U \times V \equiv U^{a} V^b h_{ab} ~.\label{innerproducts}
\ee
Clearly only the first of these can be used to define a norm. The degenerate nature of the metric means that the norm is sensitive only to the $x$ component of a vector. The second of these can be used to define angles in this geometry.

The existence of these invariant tensors is directly related to the existence of a preferred axis in the classical geometry associated to WCFTs. This axis is just given by the $t$-axis defined by the equation $x=0$. It is trivial to see that the loci of points on this axis correspond to fixed points of the transformation (\ref{eq:boost}). Notice that there is no canonical way to raise the index in $q_a$ to make another preferred vector. This means that classically, this is the only preferred axis for a WCFT. 

Quantum mechanically, the situation changes dramatically. The boost symmetry (\ref{eq:boost}) becomes anomalous as a consequence of a non-zero level for the $U(1)$ Kac-Moody algebra. This is already manifest in the anomalous transformation of the partition function under boosts (\ref{spectral}). There is a quite transparent way to see this is the case. Let us covariantize the Virasoro-Kac-Moody algebra (\ref{eq:canonicalgebra}) by defining generators:
\be\label{eq:vircov1}
J_{a, n} = \left( \begin{array}{cc}   L_n  & P_n  \end{array}\right)~ .
\ee
Using the $J$'s and \eqref{invvec}, we can write the Virasoro-Kac-Moody algebra in the compact form
\begin{eqnarray}\label{eq:vircov2}
[ J_{a , n} , J_{b , n'}]  &=&  \frac{n-n'}{2} \left( q_a J_{b , n+n'} + q_b J_{a , n+n'} \right) -  \frac{n+n'}{2} \left( q_a J_{b , n+n'} - q_b J_{a , n+n'} \right)\nonumber\\
& & + \frac{c}{12} \, g_{ab} \,  n(n^2-1) \, \delta_{n+n'} + \frac{k}{2} \, \bar{q}_a \bar{q}_b\,  n \, \delta_{n+n'}~ .
\end{eqnarray}
The classical part of the algebra can be easily written with the help of the invariant form $q_a$. Then there are two anomalous terms given by the second line of \eqref{eq:vircov2}. The term accompanying the central charge $c$ is given by an invariant tensor $g_{ab}$. However, this is not the case for the Kac-Moody anomaly $k$: here we must introduce a new one form $\bar q_a$ (normalized as $\bar q^a \bar q_a =1$) thus breaking the boost symmetry. This breaking is not severe, since it is governed by a well established anomaly; here is where the power of WCFTs reside. 

More explicitly, we choose the one-form as 
\be\label{eq:bq1}
\bar{q}_a =\left( \begin{array}{cc} 0 & 1 \end{array}\right) ~,
\ee
and its existence allows us to unambiguously define a new preferred vector $q^a$ as
\be\label{eq:qbq2}
q^a q_a =1 \quad \quad \textrm{and} \quad \quad q^a \bar q_a =0 \,.
\ee
Therefore, a WCFT has two preferred axes dictated by  $(q^a,\bar q^a)$: a classical one (the $t$ axis), and another one selected by anomalies, (which in our coordinates is given by the $x$ axis). This will be of crucial importance in what follows.

Looking ahead, and to make contact with \cite{Hofman:2014loa}, we could infer  the existence of the second preferred axis by demanding the possibility of coupling a warped quantum field theory with a scaling symmetry $ x\rightarrow \lambda x$, to geometry. There is an extra geometric structure needed, in order to have this coupling, which is nothing else but the existence of a covariantly constant vector $q^a$. With this vector one can define the tensor\footnote{Not to be confused with the currents $J_{a, n}$ defined in (\ref{eq:vircov1}).}
\be
J^a_{\phantom{a} b} = -q^a q_b ~.\label{eq:scalestru}
\ee
The eigenvalues of $J^a_{\phantom{a} b} $, which are $0$ and $-1$, select which coordinate contains a scaling symmetry and which one does not.  $J^a_{\phantom{a} b} $ is called  a {\it scaling structure} \cite{Hofman:2014loa}, in analogy to a complex structure in the usual CFT setup. As such, these preferred axes play a very similar role to that of the light cone for a CFT.

%


\section{Entanglement entropy: field theory} \label{sec:eeqft}

In this section we will compute entanglement entropy in a WCFT by using the ``Rindler method,'' i.e. via suitable coordinate maps we will show how to cast the entanglement entropy of an interval as the thermal entropy of a Rindler observer \cite{Holzhey:1994we,Casini:2011kv}.  While some technical features and outcomes of this method differ from those in a CFT$_2$, we show that the applicability of the method is equally powerful in a WCFT.

\subsection{Rindler method}

Our first task will be  to calculate the entanglement entropy of a single interval when the system is on its ground state. The background geometry is a space-time cylinder described by coordinates $(T,X)$. Following up on our previous discussion, here  $T$ is the classically $U(1)$ preferred axis and $X$ is the quantum anomaly selected axis with a scaling $SL(2,\mathbb{R})$ symmetry. In order to keep the discussion general, the identification that defines the spatial circle is given by
\be\label{eq:cyl1}
(T,X) \sim ( T + \bar{L}, X - L)~.
\ee
We will consider an interval inside this cylinder  also oriented arbitrarily
\be\label{eq:interval}
{\cal D}: \quad (T,X) \in \le[(\frac{\bar{\ell}}{2} , - \frac{\ell}{2}) , (-\frac{\bar{\ell}}{2} ,  \frac{\ell}{2}) \ri]~.
\ee
Notice that if $\frac{\bar{\ell}}{\ell} \neq \frac{\bar{L}}{L}$ then the segment is misaligned with the identification direction. For later reference we denote these two endpoints by $X_1, X_2$, where in an abuse of notation $X_1$ refers to both of the $(T,X)$ coordinates.

To quantify entanglement entropy in ${\cal D}$ we will make use of warped conformal mappings: we will show that the density matrix $\rho_{\cal D}$ describing the vacuum state on ${\cal D}$ is related via a unitary transformation to a thermal density matrix $\rho_{\cal H}$. This generalizes the results of   \cite{Holzhey:1994we,Casini:2011kv} to a case with symmetries different from that of a conformal theory, and appropriate comparisons with a CFT$_2$ will be made along the way.  To relate $\rho_{\cal D}$ to a thermal observer we  first construct a mapping from the cylinder $(T,X)$ to a set of coordinates that cover only the ``inside''  of the interval. In comparison with a relativistic system (see appendix \ref{app:cft}), we can interpret ``inside'' the interval as the causal domain of   \eqref{eq:interval}. For a warped system we are only allowed transformations of the form \eqref{app1}, and for the task at hand the appropriate transformation is
\be\label{map}
\frac{\tan \frac{\pi X}{L}}{\tan \frac{\pi \ell}{2L}} = \tanh \frac{\pi x}{\kappa}~, \qquad T+ \frac{\bar{L}}{L} X = t + \frac{\bar{\kappa}}{\kappa} x~.
\ee
We have introduced two scales, $\kappa$ and $\bar{\kappa}$, in the above map; these scales are arbitrary, and the independence of the final result on them will be used as a consistency check. In particular notice that in the $(t,x)$ coordinates the slice where the spatial identification is performed in the $(T,X)$ coordinates gets mapped to the line:
\be
t + \frac{\bar{\kappa}}{\kappa} x=0~.
\ee

 \begin{figure}
\begin{center}
\includegraphics[width=0.6\textwidth]{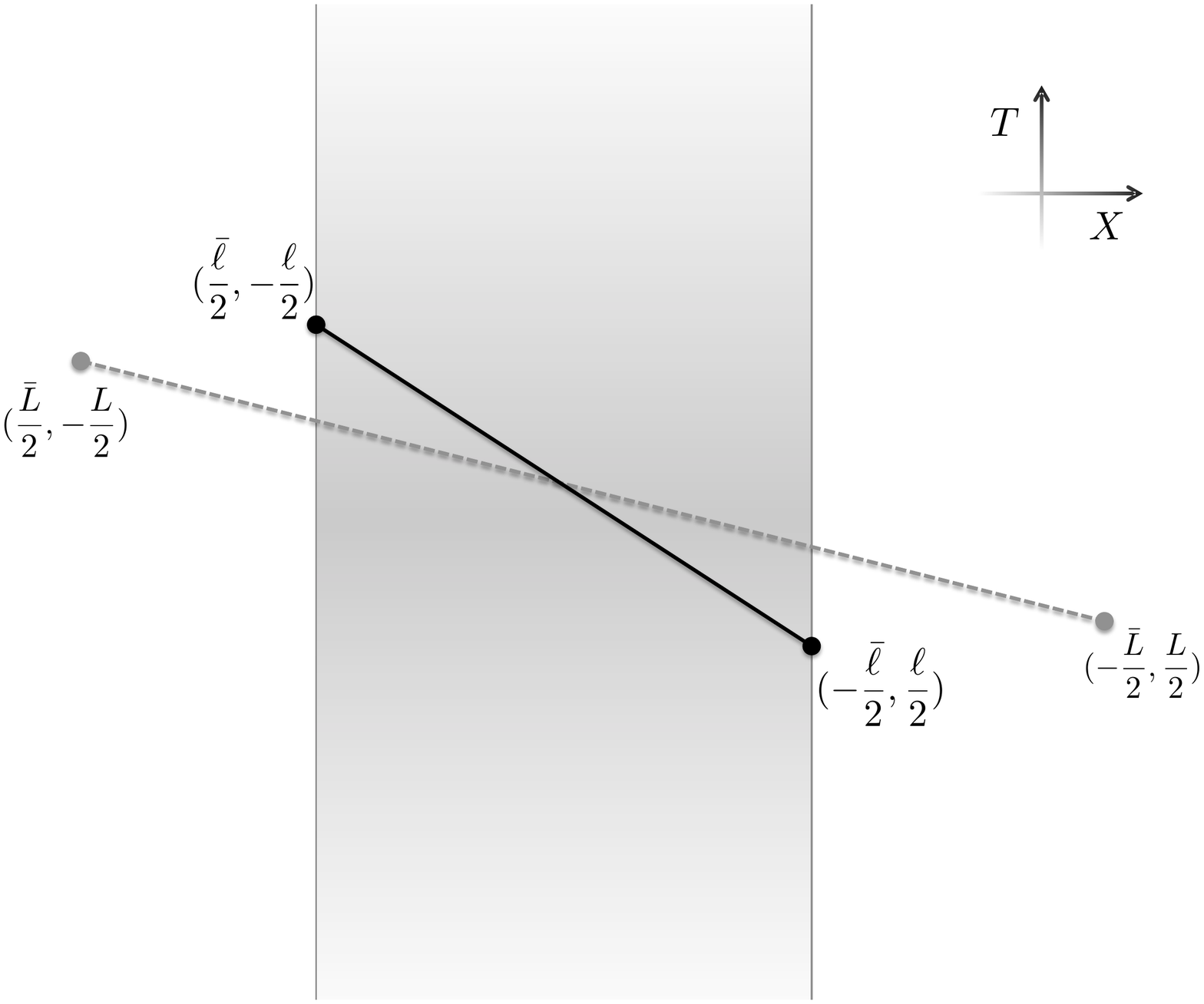}
\end{center}
\caption{ Diagram that depicts the interval ${\cal D}$ and the domain covered by the coordinates $(t,x)$ relative to $(T,X)$. Straight segment is the interval \eqref{eq:interval}; dotted line is the cylinder identification \eqref{eq:cyl1}; shaded region depicts the domain covered by $(t,x)$  on the $(T,X)$ plane.}\label{fig:wcft}
\end{figure}

This transformation has several favorable features. First, the map \eqref{map} respects the cylinder identification \eqref{eq:interval}.  The real line $-\infty < x < \infty$ covers the region $-\frac{\ell}{2} < X < \frac{\ell}{2}$ and not the rest of the cylinder. The domain of causaility, which turns out to be a strip, is depicted in figure \ref{fig:wcft}. Moreover, the expected surprise that is a direct consequence of this fact is that the map \eqref{map} induces an identification in the $(t,x)$ coordinates as:
\be\label{thermalx}
{\cal H}: \quad (t,x) \sim ( t - i \bar{\kappa}, x + i \kappa)~.
\ee
We interpret this result as the fact that the observer in $(t,x)$ coordinates perceives a thermal density matrix induced by this identification. More concretely
\be
\rho_{{\cal D}} = U \rho_{\cal H} U^{\dagger}~, \qquad \rho_{\cal H} = \exp\le(\bar \kappa P_0^{\rm cyl} - \kappa L_0^{\rm cyl}\ri)~, \label{dens}
\ee
where $U$ is a unitary transformation that implements the coordinate transformation \eqref{eq:interval}. Thus
\be\label{eq:eeth}
S_{\rm EE} = -\tr \le(\rho_{\cal D}\log \rho_{\cal D}\ri) = S_{{\rm thermal}}({\cal H})~.
\ee

For the above equality between entanglement and thermal entropy to hold, we need to be rather careful with the divergent pieces of each observable. On general grounds, we expect the entanglement entropy to have a UV divergence arising from the boundary of the interval, requiring the introduction of a short distance cutoff.  Whereas for ${\cal H}$, we expect the thermal entropy to be IR divergent due to the infinite size of $x$ in the domain of interest.  To relate these divergences, we need to obtain the length of ${\cal D}$ in the coordinate system $(t,x)$. The naive answer gives an infinite range, so we introduce a cutoff parameter $\epsilon$ which defines the new regulated interval as
\be
{\cal D}:\quad (T,X) \in \le[(\frac{\bar{\ell}}{2}- \frac{\bar{\ell}}{\ell} \epsilon \,,\, - \frac{\ell}{2} +\epsilon) \, ,\, (-\frac{\bar{\ell}}{2} + \frac{\bar{\ell}}{\ell} \epsilon \, ,\,  \frac{\ell}{2}-\epsilon) \ri]~.
\ee
Notice the factor in front of the cutoff in the $T$ direction; this is necessary to guarantee the units are correct and that the regulated interval is actually contained in the original interval. Using the map (\ref{map}) gives the image of this interval in the $(t,x)$ coordinates; we obtain
\be\label{rangex}
(t, x) \in \le[(\frac{\bar{\kappa}}{2\pi} \zeta - {\ell\over 2}\frac{\bar L}{L}  + {\bar{\ell}\over 2}\, , \, - \frac{\kappa}{2\pi} \zeta ) \, , \, (-\frac{\bar{\kappa}}{2\pi} \zeta +{\ell\over 2}\frac{\bar L}{L}  - {\bar{\ell}\over 2}\, , \,  \frac{\kappa}{2\pi} \zeta )\ri]~,
\ee
where 
\be\label{eq:cut}
\zeta = \log \left( \frac{L}{\pi \epsilon} \sin \frac{\pi \ell}{L}\right)+ O(\epsilon)~.
\ee
Notice in (\ref{rangex}) we  kept terms that are subleading relative to $\zeta$ in the small $\epsilon$ expansion; in the following we will keep these terms since they could contribute to the final answer.


\subsubsection{Entropy calculation}

Having established a relation between single interval entanglement and thermal entropy via \eqref{dens}-\eqref{eq:eeth}, we now proceed to evaluate $S_{{\rm thermal}}$. Following the notation in \eqref{eq:z1}, we denote the partition function for ${\cal H}$
\be
Z_{\bar{a}| a} (\bar \theta | \theta )~,
\ee
 where the data of the torus is built from \eqref{thermalx} and \eqref{rangex}. In other words
\be
(t,x) \sim (t+ 2\pi \bar a, x - 2\pi a) \sim (t+ 2\pi \bar \theta, x - 2\pi \theta)~,
\ee
with
\be
2\pi a = \frac{\kappa}{\pi} \zeta~, \qquad 2\pi \ba = \frac{\bar \kappa}{\pi}\zeta - \frac{\bL}{L}\ell + \bl~, \qquad 2\pi \th = - i\kappa~, \qquad 2\pi \bth = - i \bar \kappa~. \label{params}
\ee
There are two important points to emphasize at this stage. First, since the interval \eqref{rangex} is rather large we expect the edge effects to not be important and we might as well consider the identification of the interval, yielding the thermal entropy associated to a torus partition function. Second, keeping finite terms as $\zeta\to \infty$ in \eqref{params} makes the  torus non-degenerate and makes well-defined further derivations. It is not a problem as the formulae we will use yield well defined results even in the degenerate case. Still it is interesting to see that the misalignment of ${\cal D}$ with the circle identification is what breaks the degeneracy of the torus.

The problem has been reduced to that of calculating a thermal entropy.  The entropy $S_{{\rm thermal}}$ is defined as
\be\label{ent}
S_{\bar a| a} (\bar \theta | \theta) = \left(1- \theta \partial_\theta - \bar \theta \partial_{\bar \theta} \right) \log Z_{\bar a| a} (\bar \theta | \theta) ~.
\ee
It is convenient to relate this definition to the entropy associated to the partition function $\hat{Z}$ defined in \eqref{eq:defhat}; this is the frame with canonical modular properties. Using expression \eqref{spectral} and taking derivatives as in \eqref{ent} we find
\be
S_{\bar a| a} (\bar \theta | \theta)  = \hat{S} (\bar \theta - \frac{\theta}{a} \bar a\, | \, \frac{\theta}{a})~,
\ee
where we have defined
\be\label{entcan}
\hat{S} (z | \tau) = \left(1- \tau \partial_\tau - z\partial_{z} \right) \log \hat{Z} (z | \tau) ~.
\ee
This illustrates that entropy is  a robust observable for which all observers agree upon. Moreover, we can just pretend to be in the canonical circle and calculate $S_{\bar a| a} (\bar \theta | \theta)$ via  $\hat Z (z | \tau)$. And from \eqref{params}, the potentials relevant for the computation are\footnote{Another comment is in order: as promised, both $\kappa$ and $\bar \kappa$ have dropped from the computation since $\hat Z (z | \tau)$ does not dependent on them.} 
\be
\tau = - i \frac{\pi}{\zeta}~, \quad z = -  \frac{i}{2\zeta} \le(\frac{\bL}{L}\ell - \bl\ri)~.
\ee
Since $\zeta$ is divergent as  the UV cutoff $\epsilon$ is made arbitrarily small, all that is left is to evaluate $\hat Z (z| \tau)$ in the limit $\tau \rightarrow - i0$ and $\frac{z}{\tau}$ kept constant.  We can do this using Cardy-like formulae available in \cite{Detournay:2012pc,Castro:2015uaa}: from  modular transformation \eqref{eq:shz} and using the fact that the vacuum dominates the sum, the partition function is well approximated by
\be\label{eq:saddle}
\hat Z (z | \tau) = e^{i\pi \frac{k}{2} \frac{z^2}{\tau}} \hat{Z}\le(\frac{z}{\tau} \big| - \frac{1}{\tau}\ri) = e^{i\pi \frac{k}{2} \frac{z^2}{\tau}}  e^{2 \pi i \frac{z}{\tau}  P_0^{\rm vac}+ 2 \pi i \frac{1}{\tau} L_0^{\rm vac}}+\cdots~,
\ee
where $P_0^{\rm vac}$ and $L_0^{\rm vac}$ are the cylinder values of the charges in the vacuum state in the canonical circle. Notice that because the phase factor $\frac{z}{\tau}$ is constant in the limit all we need to do is to minimize $L_0$ in \eqref{eq:saddle}: for a given value of $P_0$ we expect the minimum value of $L_0$ is given by the unitarity bound 
\be
L_0^{\rm vac}= \frac{(P_0^{\rm vac})^2}{k}-\frac{c}{24}~.
\ee
If the spectrum of $P_0$ is real, as expected in a unitary WCFT we obtain\footnote{In other words, we assume that there is a state invariant under the global $SL(2,\mathbb{R})\times U(1)$ isometries of the system.}
\be
L_0^{\rm vac} = - \frac{c}{24}~, \quad P_0^{\rm vac} =0 ~. \label{normalvac}
\ee
If we allow  the spectrum of $P_0$ to be complex, which occurs often in holographic duals to WCFTs \cite{Detournay:2012pc,Compere:2013bya},  the minimum value is 
\be
P_0^{\rm vac} = -i Q~,\quad L_0^{\rm vac} = -\frac{Q^2}{k} - \frac{c}{24} ~, \label{funkyvac}
\ee
\noindent where $Q$ is a real vacuum charge.

Gathering these results, the thermal entropy of the observer ${\cal H}$ is
\be
\hat S (z | \tau) = i P_0^{\rm vac} \ell \left( \frac{\bar L}{L}  -{\bar \ell\over \ell} \right) - 4 L_0^{\rm vac} \zeta~,
\ee
where we ignored  subleading terms in $\zeta$ due to subleading corrections in \eqref{eq:saddle}. Finally, using \eqref{eq:eeth} and \eqref{eq:cut}, we find
\be\label{eq:oneint}
S_{\rm EE} = i P_0^{\rm vac} \ell \left( \frac{\bar L}{L} - \frac{ \bar \ell}{\ell}\right) -4 L_0^{\rm vac} \log \left( \frac{L}{\pi \epsilon} \sin \frac{\pi \ell}{L}\right)~.
\ee
While one might imagine that the first term is subleading it might be interesting to consider as a response of the leading value to a misalignment of the segment with respect to the circle identification. Notice it is extensive on the size of the cylinder and not periodic. As we derive this same answer using twist field correlation functions and holographically, the interpretation of these contributions will become more clear. 

\subsubsection{ Renyi entropies}

From these manipulations, it is rather straight forward to obtain Renyi entropies. These are defined as
\be
S_q = \frac{1}{1-q} \log \tr(\rho_{\cal D}^q)~. \label{renyidef}
\ee
Now the trace over (powers of) the un-normalized density matrix is computed by the following partition function:
\be
\tr \rho_{\cal D}^q = \tr_{\ba|a}\le(e^{2\pi i q\bth P_0^{c\rm yl} - 2\pi i q\th L_0^{\rm cyl}}\ri) = Z_{\ba|a}\le(q\bth | q\th\ri)~. \label{renyiZ}
\ee
Thus we want to compute
\be
S_q = \frac{1}{1-q} \log\le(\frac{Z_{\ba|a}\le(q\bth | q\th\ri)}{Z_{\ba|a}(\bth | \th)^q}\ri) ~.
\ee
Repeating again the modular manipulations as those above we find 
\be
{S_{q} = i P_0^{\rm vac} \le(\frac{\ell}{L} \bL - \bl\ri) - 2 L_0^{\rm vac}\le(\frac{1}{q} + 1\ri) \log\le(\frac{L}{\pi \ep} \sin \frac{\pi \ell}{L}\ri)}~. \label{renyians}
\ee
The $q \to 1$ limit of this agrees with \eqref{eq:oneint}. Note that the part of the entropy depending on the $U(1)$ charge does not depend on the Renyi index!  It is not evident why this is the case, and we leave further comments for the discussion. 


\subsubsection{Entanglement entropy at finite temperature}

As in the CFT case, a small tweaking of the arguments above can be used to calculate the entanglement entropy of a segment in infinite volume but at finite temperature. All we need to do is to change the map such that the original cylinder in the $(T, X)$ coordinates is identified along its thermal direction. Concretely, consider the map
\be\label{map2}
\frac{\tanh \frac{\pi X}{\beta}}{\tanh \frac{\pi \ell}{2 \beta}} = \tanh \frac{\pi x}{\kappa} ~, \quad\quad T+ \frac{\bar{\beta}}{\beta} X = t + \frac{\bar{\kappa}}{\kappa} x~.
\ee
Now the identification in the $(T,X)$ coordinates is:
\be
(T,X) \sim ( T + i \bar{\beta}, X - i \beta)~.
\ee
All the discussion goes on as before with the replacement $L \rightarrow i \beta$ and $\bar L \rightarrow i \bar \beta$. With this identifications we obtain the entanglement entropy to be:
\be
S_{\rm EE} = i P_0^{\rm vac} \ell \left( \frac{\bar \beta}{\beta} - \frac{ \bar \ell}{\ell}\right)- 4 L_0^{\rm vac} \log \left( \frac{\beta}{\pi \epsilon} \sinh \frac{\pi \ell}{\beta}\right)~.
\ee
The thermal limit is obtained by taking $\ell\to\infty$ for which $S_{\rm EE}$ reduces to the thermal entropy in the $(T,X)$ system. 

\subsection{Twist fields}\label{sec:twist}
We turn now to a slightly different interpretation of these results; one which will be useful for reproducing these results in holography, a task that we will perform in the next section. Note first that our computations up until now have been somewhat ``canonical'', in that we have been studying the problem from a Hilbert space point of view by constructing the appropriate reduced density matrix and computing its entropy. There is a complimentary ``path-integral'' point of view, in which one considers the path integral over a branched two-manifold that we will call $\sR_q$ in order to compute the Renyi entropy. The pattern of traces in the construction of the $q$-th Renyi entropy \eqref{renyidef} is implemented by considering a manifold with $q$ copies of the original space, where each replica is sewn to the consecutive one in a cyclic fashion along the interval $\sD$. We will not review this method here, and refer the unfamiliar reader to \cite{Dixon:1986qv,Knizhnik:1987xp,Calabrese:2004eu, Cardy:2007mb, Calabrese:2009qy} for a detailed discussion of this replica method applied to two-dimensional conformal field theory.  

In particular, in \cite{Calabrese:2004eu, Calabrese:2009qy} it is shown that in a conventional 2d CFT, the effect of this non-trivival topology can be implemented by considering $q$ decoupled copies of the original field theory, with the additional insertion of local {\it twist fields} $\Phi_q(X)$ at the endpoints of the interval that enforce the replica boundary conditions, coupling together the replica copies. If we denote the original theory by $\cal{C}$ and its $q$-fold copy by $\mathcal{C}^q$, then the precise statement is that for any operator $\sO(X^{(i)})$ in $\cal{C}$ located on sheet $i$ of $\sR_q$, we have
\be
\langle \sO(X^{(i)}) \rangle_{\mathcal{C}, \sR_q} = \frac{\langle \sO_i(X) \Phi_q(X_1) \Phi_q^{\dagger}(X_2) \rangle_{\mathcal{C}^q, \mathbb{C}}}{{\langle \Phi_q(X_1) \Phi_q^{\dagger}(X_2) \rangle_{\mathcal{C}^q, \mathbb{C}}}} ~,\label{twistfielddef}
\ee
where on the right hand side the expectation values are evaluated in the product theory on the ordinary complex plane $\mathbb{C}$, and $\sO_i$ denotes the operator $\sO$ belonging to the $i$-th copy. Recall that the points $X_{1,2}$ define the endpoints of the domain ${\cal D}$ in \eqref{eq:interval}.

Twist fields defined in this manner have well-defined properties under conformal transformations and can be considered to be local operators in $\mathcal{C}^q$. In this section we will study the properties of such twist fields in WCFT, determining their dimensions and $U(1)$ charges. We will not use the uniformizing map studied in \cite{Calabrese:2004eu,Calabrese:2009qy}, but will instead show that the above results for the Renyi entoropies can be re-casted in terms of twist fields. A similar method was used in \cite{Hung:2011nu} to determine the properties of twist ``surfaces'' in higher-dimensional CFTs with holographic duals.  

In this section the subregion of interest will be an interval on the plane, i.e. we will send $L, \bar{L} \to \infty$ in the spatial identification \eqref{eq:cyl1}. We will however keep track of the {\it angle} of this identification pattern $\alpha \equiv \frac{\bar{L}}{L}$. 

To identify the charges of the twist fields, we begin by computing the value of $\langle T(X) \rangle$ and $\langle J(X) \rangle$ on $\sR_q$. By the construction of $\sR_q$, this is equal to the trace of the product of the operator with the $q$-th power of $\rho_{\sD}$, i.e.
\be
\langle T(X) \rangle_{\sR_q} = \tr(T(X) \rho_{\sD}^q) ~,\qquad \langle J(w) \rangle_{\sR_q} \equiv \tr\le(J(X) \rho_{\sD}^q\ri)~.
\ee
Now $\rho_{\sD}$ is related by a unitary transformation by $U$ to the thermal density matrix $\rho_{\sH}$ in the $(t,x)$ coordinate system by \eqref{dens}. To make use of this result, we also need to understand the transformation of $T$ and $J$ under $U$. As $U$ implements the conformal transformation \eqref{map}, this is given by the anomalous transformation law \eqref{app:finite}, which we repeat here for completeness:
\begin{align}
U^{\dagger} T(X) U & = \le(\frac{\p x}{\p X}\ri)^2\le(T(x) - \frac{c}{12}\{X,x\}\ri) + \frac{\p x}{\p X}\frac{\p t}{\p X} P(x) - \frac{k}{4}\le(\frac{\p t}{\p X}\ri)^2~, \\
U^{\dagger} J(X) U & = \le(\frac{\p x}{\p X}\ri)\le(P(x) + \frac{k}{2} \frac{\p T}{\p X}\ri)~.
\end{align} 
The anomalous terms are $c$-numbers that can directly be obtained from \eqref{map}. The operator part of this expression also requires us to determine $T(x)$, $J(x)$ in the thermal state described by $\rho_{\sD}^q \equiv U \rho_{\sH}^q U^{\dagger}$. In a translationally invariant state the values of the currents are simply related to the zero modes $L_0, J_0$ by $L_0 = -a T(x)$ and $P_0 = -a J(x)$. Note also that from the definition of $Z_{\ba|a}(\btau|\tau)$ on a general torus \eqref{eq:z1} we have:
\be
\langle L_0 \rangle = -\frac{1}{2\pi i} \frac{\p}{\p\tau} \log Z_{\ba|a}(\btau|\tau)~, \qquad \langle P_0 \rangle = \frac{1}{2\pi i}\frac{\p}{\p \btau} \log Z_{\ba|a}(\btau|\tau) \ . \label{vevs}
\ee
We can now find $Z_{\ba|a}(\btau|\tau)$ in the Cardy limit by combining the Cardy result in the canonical frame \eqref{eq:saddle} with the transformation to an arbitrary frame \eqref{spectral} to find:
\be
Z_{\ba|a}(\btau|\tau) = \exp\le(\frac{i \pi k}{2} \le(\frac{\btau^2 a}{\tau}\ri) + 2 \pi i P_0^{\rm vac}\le(\frac{a \btau}{\tau} - \ba\ri) + \frac{2\pi i a}{\tau} L_0^{\rm vac}\ri) \ . 
\ee
Putting this into \eqref{vevs} we find
\begin{align}
\langle L_0 \rangle & = \frac{k}{4} \frac{\btau^2 a}{\tau^2} + P_0^{\rm vac} \le(\frac{a \btau}{\tau^2}\ri) + \frac{a}{\tau^2} L_0^{\rm vac}~, \cr
\langle P_0 \rangle & = \frac{k}{2} \frac{\btau a}{\tau} + \frac{P_0^{\rm vac} a}{\tau}~.
\end{align}
As expected for a translationally invariant state, the total value of the energy and charge scale like the length of the spatial cycle $a$. As shown in \eqref{renyiZ}, this should be evaluated on $\tau = q\theta$, $\btau = q \bth$, with $\theta, \bth$ given as before by \eqref{params}. 

Finding also the anomalous $c$-number contributions and assembling all the terms, we find after some algebra:
\begin{align}\label{vevans}
\langle T(X) \rangle_{\sR_q} & = \frac{\ell^2}{\le(X - \frac{\ell}{2}\ri)^2\le(X + \frac{\ell}{2}\ri)^2}\le(\frac{c}{24} + 
\frac{L_0^{vac}}{q^2}\ri) + \frac{i \ell \bL}{q L} \frac{P_0^{vac}}{\le(X - \frac{\ell}{2}\ri)\le(X + \frac{\ell}{2}\ri)} - \frac{k}{4}\le(\frac{\bL}{L}\ri)^2~, \cr
\langle J(X) \rangle_{\sR_q} & = \frac{\ell}{\le(X - \frac{\ell}{2}\ri)\le(X + \frac{\ell}{2}\ri)}\frac{i P_0^{\rm vac}}{q} - \frac{k}{2}\frac{\bL}{L}~.
\end{align} 
This is the desired result for the expectation values of the currents on the replica manifold\footnote{In an earlier version of this paper, there were errors in \eqref{vevans}; in particular, a term that is required for agreement with the warped CFT OPE was missing from the expression for $T(X)$. We thank G. Stettinger for bringing this to our attention \cite{StettingerThesis}.}. 

We now turn to its interpretation. From \eqref{twistfielddef}, we have:
\be
\langle T(X) \rangle_{\sR_q} = \frac{\langle T_i(X) \Phi_{q}\le(X_1\ri) \Phi^{\dagger}_{q} \le(X_2\ri)\rangle}{\langle \Phi_{q}\le(X_1 \ri) \Phi^{\dagger}_{q}\le(X_2 \ri)\rangle}~, \qquad \langle J(X) \rangle_{\sR_q} = \frac{\langle J_i(X) \Phi_{q}\le(X_1 \ri) \Phi^{\dagger}_{q} \le(X_2\ri)\rangle}{\langle \Phi_{q}\le(X_1\ri) \Phi^{\dagger}_{q}\le(X_2\ri)\rangle}~. \ee
Now, as noticed in \cite{Calabrese:2004eu, Calabrese:2009qy}, the form \eqref{vevans} for the stress tensor (and in our case the $U(1)$ current) expectation value is equivalent to the Ward identity for the conformal primary $\Phi_q(X)$.

More explicitly, the OPE of the twist field with the $U(1)$ current takes the form $J(x) \Phi_{q}(y) \sim \frac{i Q_{q} \sO}{x-y}$ (and similarly for the stress tensor). This determines the singularity structure of the correlation function. In particular, the unfamiliar subleading stress tensor singularity that is proportional to $P_0^{vac}$ in \eqref{vevans} arises from the different functional form of the two-point function of primary operators in WCFT \cite{StettingerThesis}. The functions appearing in \eqref{vevans} are the unique analytic functions of $X$ that have the correct singularity structure and approach a constant at infinity. Thus the charges may be read off from the singularities in \eqref{vevans}. We must multiply by a factor of $q$ to go from $T_i$ to the full stress tensor $T$ on $\mathcal{C}^q$, leading to the following values for the conformal dimension and charge of the twist field $\Phi_{q}$:
\be
\Delta_{q} = q\le(\frac{c}{24} + \frac{L_0^{\rm vac}}{q^2}\ri)~, \qquad Q_{q} = P_0^{\rm vac}~.
\label{unfinishedtwistdims1}
\ee
It is interesting to note that the $q \to 1$ limit of these charges is not obviously zero. We will return to this point, but first we proceed to compute the Renyi entropy itself. As usual, the two-point function of these twist operators determines the partition function on the $q$-sheeted Riemann surface. This in turn determines the Renyi entropy, and we have:
\be
S_q = \frac{1}{1-q} \log\frac{\tr \rho^q}{(\tr \rho_1)^q} \sim \frac{1}{1-q} \frac{\langle \Phi_{q}\le(X_1 \ri) \Phi_{q}^{\dagger}\le(X_2 \ri)\rangle}{\langle \Phi_{1}\le(X_1\ri)\Phi_{1}^{\dagger}\le(X_2\ri)\rangle^q}~. \label{renyicorr}
\ee
We now need to determine the two-point function of the twist field on the plane. We expect the 2-point correlation function of  primary operators on the plane to be fixed by symmetries: while this is true, the precise implementation of these symmetries in the case of WCFT is somewhat novel. 

Define $\Delta X^{a} \equiv X_2^{a} - X_1^{a}$. As we are in flat space, we need not distinguish between tangent-space and spacetime indices (equivalently, there exists a canonical vielbein $\tau^{a}_{\mu} \equiv \delta^{a}_{\mu}$). Correlation functions should now depend only on invariants associated with $\Delta X^a$. One such invariant is its norm, as defined in  \eqref{innerproducts}:
\be
\sqrt{\Delta X^a \Delta X^b g_{ab}} = \left| X_1^{x} - X_2^{x} \right| = \ell~,
\ee
where $X_{1,2}$ are given by \eqref{eq:interval} and the metric is defined in \eqref{2dtensors}. By inspection, it is clear that there does not appear to be an invariant built from $(\Delta X^a, g_{ab},h_{ab})$ that is sensitive to the separation $\bar{\ell}$ in the time direction. This is not surprising since the separation along the $T$-axis  is not invariant under the boost $T \to T + v X$, and hence we don't expect $\bar\ell$ by itself to be a good measure. However, there is another invariant that we can be build by introducing another vector: denote by $V^a$ the vector that corresponds to the identification pattern associated to the cylinder \eqref{eq:cyl1}
\be
V^a \equiv \le(\begin{tabular}{c} $L$ \\ ${\overline L}$ \end{tabular}\ri)~.
\ee
We may now define a normalized vector
\be
 n^a \equiv \frac{V^a}{\sqrt{V^a V^b g_{ab}}} ~,
 \ee
 which remains finite as we take the cylinder very large, carrying only the information of the {\it angle} of the identification pattern. Now, the cross product of $n^a$ with $\Delta X^a$ is given by
\be
s \equiv n^a \Delta X^b h_{ab} = \bar{\ell} - \ell \frac{\overline L}{L} ~,
\ee
and it is also a boost invariant. We see that $s$ is a covariantized measure of the separation in the time direction. At this point the correlator is an arbitrary function of $s$ and $\ell$. Now the operator has conformal dimension $\Delta_q$ with respect to scalings of the $X$ direction, as measured by $\ell$. It also has a $U(1)$ charge $Q_q$ with respect to ``translations'' in the $T$ direction, as measured by $s$. 
Thus the correlation function takes the form\footnote{The $U(1)$ direction is anomalous, and hence \eqref{eq:coorphi} comes with a few caveats. We are assuming implicitly that vacuum state used to compute the expectation value in \eqref{eq:coorphi} is neutral under the $U(1)$ charge; this implies that the path integral will only depend on invariant quantities. See Section 3.3 of \cite{Castro:2015uaa} for the analogous arguments for the partition function. If the vacuum state is charged, then the extra terms due to the anomaly are simple to quantify by keeping track of the anomalous transformation of the path integral. }  
\be\label{eq:coorphi}
\langle \Phi_{q}\le(X_1 \ri) \Phi_{q}^{\dagger}\le(X_2 \ri)\rangle \sim {\ell}^{-2\Delta_{q}} \exp\le(-i s Q_{q}\ri) \ . 
\ee
Putting in the values of \eqref{chargeans} and evaluating \eqref{renyicorr} we reproduce the previous value for the Renyi entropy \eqref{renyians}, as expected.

We now discuss some interesting features of the twist fields defined above. For example, consider first the case $q = 1$; in this case we have not traced anything out and are simply considering the expectation value of the stress tensor on the plane. One might then expect the stress tensor to be non-singular everywhere. Instead, however, we find a nontrivial answer with:
\be
\Delta_{1} = \frac{c}{24} + L_0^{\rm vac}~, \qquad Q_{1} = P_0^{\rm vac}~.
\ee
In the usual case, we study a vacuum that is $SL(2,\mathbb{R}) \times U(1)$ invariant, as in \eqref{normalvac}: then we have $L_0^{\rm vac} = -\frac{c}{24}$ and $P_0^{\rm vac} = 0$, and both of the charges above vanish, resulting in a regular stress tensor. 

However, if the vacuum is {\it not} invariant under $SL(2,\mathbb{R}) \times U(1)$ -- as in \eqref{funkyvac} -- then the 
vacuum on the cylinder maps to a non-trivial operator on the plane. This vacuum operator may be understood as being $\Phi_{1}$. Any computation performed on the plane will involve an insertion of $\Phi_{1}$ that ``creates the vacuum'', as well as a corresponding insertion of $\Phi_{1}^{\dagger}$ to ``annihilate the vacuum''. The freedom to move these operators around means that there is no translationally invariant quantization of this theory on the plane. 

In our precise computation, these insertions of the vacuum operator have localized at the endpoints of the interval. In the $q$-fold theory we obtain $q$ copies of $\Phi_{1}$. We might attempt to separate this vacuum contribution from the twist field by subtracting $q$ times its contribution to obtain the charges of the twist field itself:
\be
\Delta^{\rm twist}_q = L_0^{\rm vac}\le(\frac{1}{q} - q\ri)~, \qquad Q^{\rm twist}_{q} = P_0^{\rm vac}\le(1 - q\ri) ~.\label{chargeans}
\ee
This subtraction -- while conceptually useful -- is somewhat heuristic, and cannot really be justified unless there is some other principle (e.g. a large central charge and gap in the spectrum) that allows us to add conformal dimensions. It is thus reassuring that in the actual computation of the correctly normalized Renyi entropy, this subtraction happens {\it automatically} between the numerator and denominator of \eqref{renyicorr}.   

A consequence of this is that the entanglement entropy is determined by the value of the vacuum charges $L_0^{\rm vac}$ and $P_0^{\rm vac}$, not by the central charge. As we have stressed above, if the vacuum is not invariant under the appropriate conformal group, these are not directly correlated. Another situation with a similar mismatch is Liouville theory, where it is well-known that the Cardy limit of the {\it thermodynamic} entropy is also controlled by the vacuum charges and not the central charge of the theory \cite{Seiberg:1990eb,Kutasov:1990sv,Carlip:1998qw}, and our discussion above can be viewed as an extension of those results to the entanglement entropy. A similar result has been obtained in the context of non-unitary CFTs in \cite{Bianchini:2014uta}. 


\section{Entanglement entropy: holography in geometric language} \label{sec:holgeom}

The results of the previous section show clearly that the symmetries of the problem are enough to determine completely the entanglement entropy of a single interval when the system is in its ground state and the interval has an arbitrary orientation with respect to the identifications of a space-time cylinder where our theory is defined. As such we expect that any correct holographic description of these systems will share the same property. We will show that this is the case by using a geometric description of holographic duals of WCFT; in the following section we will derive these results using a Chern-Simons formulation of the holographic dual. 

It is well known that for standard CFTs the way to obtain entanglement entropies from holography is to perform a calculation of the minimal area for a bulk surface attached to the edge of the entanglement region at the boundary. This is nothing else than the Ryu-Takayanagi formula \cite{Ryu:2006bv,Ryu:2006ef}. This prescription takes a special form for CFTs in two space-time dimensions. In that case, the minimal surface corresponds to a geodesic in the bulk describing the trajectory of a semi-classical particle. This suggests a connection to the field-theory computation involving twist fields: the two-point function of twist fields is related to the entanglement entropy, and the geodesic in the bulk is known to compute boundary theory two-point functions for operators with large conformal dimensions. The only necessary data to perform the computation is the quantum numbers of the twist fields. Their quantum numbers are fixed completely by the charges of the vacuum state, which are in turned determined by anomalies (under some assumptions). Plugging this data into the geodesic calculation in the holographic bulk yields the Ryu-Takayanagi formula, including the correct factors of $\frac{1}{4 G_N}$. The preceding discussion is heuristic, essentially because one can really understand the twist field as a probe of a fixed background only in the limit that the Renyi index used in the replica trick is taken $q \to 1$. 

Under certain circumstances, however, it can be made precise through a careful implementation of this limit in the bulk \cite{Lewkowycz:2013nqa}. We will not attempt to do so here. Instead, given that we have a good understanding of the properties of twist operators for WCFTs, as described in Section \ref{sec:twist}, we will assume that the line of reasoning described above is valid, and simply calculate the appropriate two-point functions through semi-classical particle trajectories in the bulk, relating them at the end to entanglement entropy. What this means is that we will consider background and dynamic fields that are fully gauge invariant under tangent space gauge transformations. This point of view will make manifest the comparison with standard geometric concepts like that of geodesics in Riemannian geometry versus warped geometry.


\subsection{A geometric background}

In order to understand how to describe particle dynamics in warped geometry, it is first important to explain what are the necessary structures to describe the background geometry. Let us remind the reader that in order to describe semi-classical particle dynamics it is not necessary to consider a dynamical geometry. All we really need are fixed background fields. In section \ref{subsec:anom}, a brief outline of the flat (i.e. tangent space) geometry that couples to WCFT was given. It was argued that it can be constructed out of classically invariant tensors $\bar q^a$ and $q_a$ as well as out of the preferred tensors $q^a$ and $\bar q_a$ selected by quantum anomalies. In  the following we will briefly elaborate on this formulation for $(d+1)$-dimensional warped geometry; for a complete discussion on this subject we refer the reader to \cite{Hofman:2014loa}.

To discuss warped geometry in dimensions larger than $(1+1)$, the first step is to extend the number of coordinates as
\be
\left(\begin{array}{c} x \\ t\end{array}\right) \quad \longrightarrow \quad \left(\begin{array}{c} x^I \\ t\end{array}\right)~, 
\ee
\noindent where the $I=1, \ldots , d$ spans usual coordinates transforming under an $SO(d)$ symmetry. Therefore, the relevant tensors are extended as:
\be
q_a \rightarrow q_a^I~, \quad\quad q^a \rightarrow q^{a I} ~.
\ee
As expected there now also exists a $SO(d)$ invariant tensor $\delta_{I J}$.\footnote{Here we consider a purely spatial (i.e. Euclidean) extension of the $x$ coordinate, and $t$ remains the preferred $U(1)$ axis.} The vectors $(\bar q_a, \bar q^a)$ are still of the form \eqref{invvec} and \eqref{eq:bq1} extended in the obvious way to $(d+1)$ dimensions; the generalizations of \eqref{2dtensors} and \eqref{eq:qbq2} are
\be
g_{ab}=\delta_{IJ}q_a^I q_b^J~, \quad q^{a I} q_{a J} =\delta^I_{J} ~,\quad q^{aI} \bar q_a=0~, \quad \bar q^{a} \bar q_a=1~.
\ee

Now we would like to extend these notions to curved space. In a nutshell, all we need to do is add vielbein fields that map the vector space in the base manifold to tangent space. Let us call these invertible fields $\tau^a_\mu$. Using these fields we can build spacetime tensors as in standard geometry. Lower index tensors built from these objects are
\be\label{eq:GA}
G_{\mu \nu} = \delta_{I J} q_a^I q_b^J \tau^a_\mu \tau^b_\nu~, \quad\quad \bar A_\mu = \bar q_a \tau^a_\mu~.
\ee
We define upper index tensors as
\be
G^{\mu \nu} = \delta_{I J} q^{a I} q^{b J}  \tau_a^\mu \tau_b^\nu ~,\quad\quad \bar A^\mu = \bar q^a \tau_a^\mu~.
\ee
Notice that the orthogonality properties of the $(q^I,\bar q)$ vectors imply that 
\be
G_{\mu \nu} \bar A^\mu = G^{\mu \nu} \bar A_\mu = 0 ~,\quad G^{\mu \nu} G_{\nu \rho} = \delta^\mu_\rho - \bar A^\mu \bar A_\nu~,  \quad \bar A^\mu \bar A_\mu =1~,
\ee
\noindent which shows that the $G_{\mu \nu}$ metric is degenerate. 

This is all the geometric structure needed (and available) to describe the trajectory of a semi-classical particle. A complete discussion of the fully dynamical bulk theory, called Lower Spin Gravity \cite{Hofman:2014loa},  should include as well dynamics for the geometrical variables $G$ and $\bar A$. For our immediate purpose of evaluating holographic entanglement entropy, we only need the values of these background fields which correspond to the vacuum state of the dual WCFT.  

In $(2+1)$-dimensions, lower spin gravity admits a description as a $SL(2,\mathbb{R}) \times U(1)$ Chern-Simons theory \cite{Hofman:2014loa}. A consequence of this is that $G_{\mu \nu}$ must describe a $SL(2,\mathbb{R})$ invariant geometry while $\bar A_\mu$ must be a flat $U(1)$ connection deformed by a killing vector of the $SL(2,\mathbb{R})$ invariant metric. This freedom in the deformation is completely analogous to the freedom of selecting a particular vielbein from a Chern-Simons connection in higher spin setups. In relation to the standard warped AdS$_3$ setup \cite{Anninos:2008fx}, this deformation corresponds to the warping parameter. We will see below that the value of this deformation has no physical consequence in our setup.

By implementing the above features, the $(2+1)$-dimensional background geometry for warped holography is
\be\label{eq:back}
G_{\mu\nu}dx^\mu dx^\nu = R^2 \, \frac{dr^2 + dX^2}{r^2} ~, \quad\quad \bar A_\mu dx^\mu = dT + \beta dX + \gamma \frac{dX}{r} ~.
\ee
Notice that the $SL(2,\mathbb{R})$ invariant geometry is nothing else than an euclidean $AdS_2$ subspace of our warped geometry. $R$ is the AdS$_2$ radius; $\beta$ parameterizes the flat $U(1)$ connection while $\gamma$ does the equivalent for the killing vector deformation. 

The $X$ and $T$ coordinates \eqref{eq:back} parameterize the boundary of our bulk geometry, and in the following we will pick the topology of the boundary to be a cylinder. In particular, we impose the identification
\be
(T, X ) \sim (T + \bar L, X -L)
\ee
\noindent as in our field theory computation (\ref{eq:cyl1}). If the geometry is regular (smooth) in the interior we must impose the vanishing of the $\bar A$ holonomy at the center $r \rightarrow \infty$ of Euclidean AdS$_2$ over this cycle:
\be\label{betaid}
\int \bar A = \bar L + \beta L = 0 \quad \rightarrow \beta = -\frac{\bar{L}}{L}  ~.
\ee
\noindent This fixes the value of $\beta$ for our background field $\bar A$.

\subsection{Worldline action}

We now have all ingredients to  describe the coupling of a point particle to the background geometry $( G_{\mu \nu}, \bar A_\mu )$. 
It does not take too much work to write down the most general fully covariant action to lowest non trivial order on the trajectory field $x^\mu(\tau)$:
\be\label{action1}
S =  \frac{1}{2} \int d\tau e^{-1} \dot x^\mu G_{\mu \nu} \dot x^\nu + \frac{m^2}{2} \int  d\tau \,e + h \int d\tau \bar{A}_\mu \dot x^\mu ~,
\ee
\noindent where we introduced a worldline einbein $e$ to make sure the action is invariant under worldline reparameterizations. Notice that while we can redefine $x^\mu$ to set the normalization of the first term to be canonical, the worldine action possess two physically meaningful constants, as expected: $m^2$ and $h$. Because a particle has to transform in a representation of the underlying $SL(2,\mathbb{R}) \times U(1)$ symmetry we expect it to be defined by two quantum numbers: a $SL(2,\mathbb{R})$ Casimir, and a $U(1)$ charge. 

The equations of motion for this action, obtained by varying with respect to $e$ and $x^\mu$, can be written in a compact form:
\begin{eqnarray}
\dot x^\mu G_{\mu \nu} \dot x^\nu &=& m^2 e^2 ~,\\
G_{\mu\nu} \frac{d}{d\tau} \left(e^{-1} \dot x^\nu\right) +\Gamma_{\mu \alpha \beta}  \left(e^{-1} \dot x^\alpha \right) \dot x^\beta  &=& h \dot x^\nu T_{[\mu \nu]}~,
\end{eqnarray}
\noindent where we have defined an affine connection
\be
\Gamma_{\mu \alpha \beta} = \frac{1}{2} \left[ \partial_{(\alpha} G_{\beta)\mu} - \partial_\mu G_{\alpha \beta} \right]~,
\ee
and a torsion field
\be
T_{[\mu \nu]} = \partial_{[\mu} \bar A_{\nu]} \,.
\ee

We can obtain a standard looking geodesic equation (corrected by torsion) by picking a preferred time parameterization given by the gauge choice $e=m^{-1}$:
\begin{eqnarray}
\dot x^\mu G_{\mu \nu} \dot x^\nu &=& 1~, \\
 G_{\mu\nu} \ddot x^\nu+\Gamma_{\mu \alpha \beta}  \dot x^\alpha \dot x^\beta  &=& \frac{h}{m} \dot x^\nu T_{[\mu \nu]} \, . \label{eq:geo1}
\end{eqnarray}

This is the geodesic equation in warped geometry. It is not universal, i.e. it depends on a parameter $\frac{h}{m}$, just as the geodesic equation for a normal charged particle depends on $\frac{q}{m}$.  Notice the following peculiarity: this equation is first order for one of the components of the trajectory, as can be seen by multiplying by $\bar A^\mu$. We will see in the next section that this fact has important consequences. For our backgrounds of interest (\ref{eq:back}), this component becomes arbitrary and all paths that have the appropriate boundary conditions will satisfy the geodesic equation (\ref{eq:geo1}). This is directly related to the Chern-Simons origin of our theory.

\subsection{Two-point functions for holographic WCFTs}

Now that we have the necessary particle action we can calculate the two point function of a heavy operator in our WCFT by calculating the Euclidean on shell action with the appropriate boundary conditions fixing the trajectory to the boundary of our three dimensional geometry at the point where the operators are inserted. More concretely we will evaluate
\be\label{eq:two}
\langle \Phi(X_1,T_1) \Phi(X_2,T_2) \rangle \sim e^{-S_E^{\rm on-shell}[\Delta X,\Delta T]} ~,
\ee
\noindent where $\Delta X = X_1 -X_2$ and $\Delta T = T_1 - T_2$. The Euclidean action is obtained by Wick rotating the time component of our geometry. In \eqref{eq:back} we have picked euclidean $AdS_2$ for our $SL(2,\mathbb{R})$ invariant metric, hence it makes sense to consider the direction singled out by $\bar A^\mu$ to be the time direction. This amounts to considering the following euclidean action:
\be\label{actioneuc}
S_E =  \frac{1}{2} \int d\tau e^{-1} \dot x^\mu G_{\mu \nu} \dot x^\nu + \frac{m^2}{2} \int  d\tau\, e + i h \int d\tau \bar{A}_\mu \dot x^\mu \, .
\ee

We are interested in finding solutions to the equations of motion in \eqref{actioneuc} when the background is given by \eqref{eq:back}-\eqref{betaid}. We could manipulate explicitly \eqref{eq:geo1}, however it is useful to exploit certain symmetries of the background. Since there is no explicit $T$ and $X$ dependence in \eqref{eq:back}, and hence \eqref{actioneuc}, the canonical momenta $(P_T, P_X)$ are conserved. From varying \eqref{actioneuc} we get 
\be\label{eq:ptpx}
P_T = \frac{\delta S_E}{\delta \dot T}=i h ~, \quad \quad P_X = \frac{\delta S_E}{\delta \dot X}= e^{-1} \frac{\dot X}{r^2} R^2 + i h \beta+ i h \frac{\gamma}{r} ~.
\ee
The canonical momentum to $r(\tau)$, which is not conserved, is given by   
\be\label{Pr}
P_r^2=\frac{1}{r^2} \left[ m^2 R^2 - r^2 (P_X - i h \beta - i h \frac{\gamma}{r})^2\right] ~,
\ee
where we used the constraint coming from the variation of $e$. The main appeal of writing these momenta is that the on shell action is given by 
\be
S_E^{\rm on-shell} = \int P_\mu dx^\mu ~,
\ee
which is the usual expression for systems satisfying a Hamiltonian constraint. Since $(P_T, P_X)$ are constant, we have 
\be
S_E^{\rm on-shell} = ih \Delta T + P_X \Delta X + 2 \int_0^{r_c} P_r dr ~,
\ee
\noindent where $r_c$ is the critical turning point of the trajectory. Here we have made a choice: we consider only trajectories that start and end at the boundary $r\to 0$ and with non-trivial separation along $X$ and $T$. 

There are two constants left to determine in terms of our boundary conditions: $r_c$ and $P_X$. First, as in any projectile motion, the turning point is defined by the vanishing of $P_r$. From (\ref{Pr}) we find
\be\label{eq:rc}
r_c = \frac{\pm m R + i h \gamma}{P_X- i h \beta} \, .
\ee
Second, we need to relate $P_X$ to the distance traveled along the $X$ direction. Using \eqref{eq:ptpx} gives
\bea\label{eq:pxx}
\Delta X = 2 \int_0^{r_c} \frac{\dot X}{\dot r} dr &=& 2 \int_0^{r_c}  \frac{r}{R} \frac{P_X- i h \beta - i h \frac{\gamma}{r}}{\sqrt{m^2 - \frac{r^2}{R^2}(P_X- i h \beta - i h \frac{\gamma}{r})^2} }dr \cr &=& 2 \frac{\sqrt{m^2 R^2 + h^2 \gamma^2}}{P_X - i h \beta} ~.
\eea
This fixes $P_X$. 

Our action can now be written as 
\be
S_E^{\rm on-shell} = i h \Delta T +P_X\Delta X + 2 \int_0^{r_c} {\sqrt{m^2 R^2 -r^2 (P_X  - i h \beta- i h \frac{\gamma}{r} )^2}} \, {dr\over r} ~. 
\ee
The integral diverges near the endpoints of the curve $r\to 0$, and to evaluate  $S_E^{\rm on-shell}$ we need to introduce a cutoff: take the endpoints to  lie at $r=\Lambda$, and as we take $\Lambda\to 0$ we only keep the leading terms in $\Lambda^{-1}$ that involve $\Delta X$ or $\Delta T$. Implementing this cutoff gives
\be\label{eq:son}
S_E^{\rm on-shell} \sim i h \Delta T + i h \beta \Delta X+ 2\sqrt{R^2 m^2 + h^2 \gamma^2} \log{\frac{\Delta X}{\Lambda}} ~,
\ee
where we also used \eqref{eq:rc} and \eqref{eq:pxx}.

From here we can estimate the (normalized) two point function; using  \eqref{eq:two} and \eqref{eq:son} gives
\be\label{2p}
\langle \Phi(X_1,T_1) \Phi(X_2,T_2)\rangle \sim e^{-S_E^{on shell}} =e^{i h \left( \frac{\bar L}{L}\Delta X- \Delta T\right) } \frac{1}{\left(\Delta X\right)^{ 2\sqrt{R^2 m^2 + h^2 \gamma^2}}}  ~.
\ee
In this expression we used that  the value of $\beta$ is given by \eqref{betaid}, which is obtained as a regularity condition in the bulk. 

The two point function \eqref{2p} agrees exactly with the expected result from field theory (\ref{eq:coorphi}). Moreover we can relate the charge and scaling dimension of our WCFT field  with the mass and $U(1)$ charge of the bulk particle; by comparing both expressions we obtain
\be
Q_\Phi = h ~, \quad\quad \Delta_\Phi = \sqrt{R^2 m^2 + h^2 \gamma^2} ~.
\ee
Notice that the values of $R$ and $\gamma$, which are background parameters do not affect the operator properties as $m$ can be chosen to adjust them to any desired value. As part of the holographic dictionary, we would relate the central charge $c$ to $R$ and the $U(1)$ anomaly $k$ to $\gamma$. Moreover, if $\gamma$ is real we can always re-normalize it away 
\be
h \rightarrow \frac{h}{\gamma} ~. \quad \quad \Delta T \rightarrow \gamma \Delta T ~,
\ee
and hence set $\gamma=1$. This is the expected transformation rule for the  $U(1)$ anomaly $k$. Finally, in order to reproduce the results for entanglement entropy all we need to do is to evaluate the two point functions above at the values $Q_\Phi$ and $\Delta_\Phi$ obtained for the WCFT twist operators (\ref{unfinishedtwistdims1}).

At this point it is important to point out  that the equations of motion do not fix the trajectory in the $T$ coordinate. Evolution in this direction is not only undetermined, it is not important: the only important data are the endpoints along this axis. This is what we expect from a $U(1)$ Chern-Simons  theory. This point could be unsettling. However, this feature is what makes the calculation convergent. In usual Warped AdS holography, the local Riemannian description of the geometries introduces in our notation  the following term to the action of a point particle 
\be
\Delta S = \frac{\alpha}{2} \int d\tau e^{-1} \left(\bar{A}_\mu \dot x^\mu\right)^2 ~.
\ee
due to using $G_{\mu \nu} +\alpha \bar{A}_\mu \bar{A}_\nu$ as an effective geometry. This term gives dynamics to the time component. However,  it does also leads either to further divergences \cite{Anninos:2013nja} or it can also induce a scaling dimensions of the field that depends on the value of $\Delta T$ \cite{Faulkner:2009wj}. This is the situation in the standard setup in AdS$_2$ holography. These are the usual problems with holographic renormalization in warped spaces. Lower spin gravity avoids this problem elegantly by suppressing this interaction term from the action.


\section{Entanglement entropy: holography in Chern-Simons languange } \label{sec:holCS}

In \cite{Hofman:2014loa} it was shown that the minimal way to describe the holography of warped conformal field theory was in terms of a $\slt \times U(1)$ Chern-Simons theory in the bulk. The relevant bulk degrees of freedom are a $\slt$ gauge field $B$ and a $U(1)$ gauge field $\bB$, and the bulk action is simply
\be
S = k_{CS} \int \Tr\le(B \wedge dB + \frac{2}{3} B \wedge B \wedge B\ri) - \xi \int \bB \wedge d \bB ~.
\ee
Here $k_{CS}$ can be related to the central charge. $\xi$ is a parameter whose value (but not its sign) can be changed by real rescalings of $\bB$, and thus can be set to one of $\pm 1, 0$. In what follows we will use an explicit matrix realization of the $\slt$ algebra, which is given by\footnote{Of course, our results don't rely on this choice of representation.} 
\be\label{eq:convsl2}
L_{1}=
\left(
\begin{array}{cc}
0 & 0\\
-1 &0
\end{array}
\right)~,
\quad \quad 
L_{-1}=
\left(
\begin{array}{cc}
0&1\\
0&0
\end{array}
\right)~,
\quad \quad
L_0=
{1\over 2} \left(
\begin{array}{cc}
     1&0\\
     0&-1
\end{array}
\right)~.
\ee

The equations of motion simply tell us that both gauge connections are flat. The vacuum \eqref{eq:back} corresponds to the flat connections
\be
B = L_0 d\rho + e^{\rho} L_+ dX ~, \qquad \bB = dT + \beta dX ~. \label{ads2vac} 
\ee
The topology of the 3 manifold has a contractible cycle described by the identifications $(T, X ) \sim (T + \bar L, X -L)$. If this configuration is smooth,  the holonomy of $\bB$ along this cycle must be trivial which sets $\beta = -\frac{\bar{L}}{L}$. 

To connect this Chern-Simons formalism to the geometric language of the previous section, we must pick a two-dimensional subspace of $\slt$ to associate with the two scaling directions in the bulk. We can then project the normal Killing form of $T_{mn}$ of $\slt$ down onto this two-dimensional subspace to obtain a degenerate Killing form $\hat{T}_{mn}$

\be
\hat{T}_{mn} = T_{m n } - \zeta_m \zeta_n ~,
\ee
\noindent where $\zeta_m$ is the Killing vector of choice that was projected out.
This degenerate Killing form is used to find the geometric degenerate metric defined in the previous section:
\be
G_{\mu\nu} = \hat{T}_{mn} B^m_{\mu} B^n_{\nu}~, \label{metident}
\ee
The conjugacy class of the omitted generator determines the signature of the metric $G_{\mu\nu}$. For example, if we take the subspace orthogonal to the hyperbolic generator $L_+ + L_-$, then we obtain from \eqref{ads2vac} the Euclidean signature metric
\be
G_{\mu\nu} dx^{\mu} dx^{\nu} = \ha(d\rho^2 + e^{\rho} dX^2)~. \label{eucBulk}
\ee
Taking instead the subspace orthogonal to the elliptic generator $L_+ - L_-$ would make $X$ into a timelike coordinate. These considerations will turn out to be important when determining boundary conditions on our probe. 

We would now like to obtain the results described in the previous sections --regarding entanglement entropy and correlation functions-- from the Chern-Simons description. This boils down to coupling massive particles to Lower Spin Gravity using Chern-Simons variables. 

A version of this problem has been studied previously in \cite{Ammon:2013hba} for $\slt \times \slt$ gravity (as well as a higher-spin generalization), where it was shown that bulk Wilson lines in an infinite-dimensional highest-weight representation of $\slt \times \slt$ capture the physics of heavy particles in the bulk. The Casimirs characterizing the representation are related to the mass and other charges of the particle. To compute a boundary theory correlation function the Wilson lines are picked to intersect the boundary. In this section we will adapt that discussion to Lower Spin $\slt \times U(1)$ gravity.  We first briefly review the prescription of \cite{Ammon:2013hba}, referring the reader to that work for a more detailed discussion. 

To compute a Wilson line in an infinite-dimensional representation of the gauge group, we first construct an auxiliary quantum mechanical system living on the worldline. This quantum mechanical system is picked to have a global symmetry group such that its Hilbert space furnishes precisely the infinite-dimensional representation in question. We then couple this auxiliary system to the bulk gauge fields (viewed as external sources for the global symmetry along the worldline) in the standard way. Integrating out this auxiliary system then computes the Wilson line in question. 

In the case where we have two $\slt$ gauge fields (as is appropriate for standard AdS$_3$ gravity), the correct quantum mechanical system is a particle living on the $\slt$ group manifold, $U \in \slt$. Two copies of $\slt$ act naturally from the left and right as
\be 
U \to L U R \qquad L, R \in \slt \ . \label{Urot}
\ee
 It can be shown that upon quantization the Hilbert space of a particle moving on $U$ transforms as a highest weight representation under both $\slt$'s \cite{Dzhordzhadze:1994np}.  Note that the group manifold $\slt$ is actually itself AdS$_3$, and thus we are simply re-asserting the familiar fact that single-particle states on AdS$_3$ transform as highest-weight representations under its isometry group (see e.g. \cite{Maldacena:1998bw}). 

Now the worldline action describing the system was shown to be
\be
S[U,P,\lam;A,\bar A] = \int_C ds \le( \Tr\le(P U^{-1} \frac{DU}{ds}\ri) + \lam \le(\Tr(P^2) - c_2\ri)\ri) ~,
\ee
where $P$ is the momentum conjugate to $U$, $\lam$ is a Lagrange multiplier that guarantees that the representation has quadratic Casimir equal to $c_2$ (which can be related to the mass of the particle) and the covariant derivative is
\be
\frac{DU}{ds} =  \frac{dU}{ds} + A_s U - U \bar  A_s~, 
\ee
where the external sources $A_s, \bar A_s$ denote the pullback of the bulk gauge field to the path $X^{\mu}(s)$ via $A_s \equiv A_{\mu} \frac{dX^{\mu}}{ds}$. We may now compute the Wilson line by performing the path integral over all worldline fields,
\be
W(A, \bar A) = \int [\sD U \sD P \sD \lam] \exp(-S[U,P,\lam; A,\bar A]) ~,
\ee
In the semi-classical limit this amounts to simply computing the bulk on-shell action. Techniques -- which we will review below -- were developed in \cite{Ammon:2013hba} to compute this action purely algebraically in terms of data characterizing the flat connections $A, \bar A$.\footnote{In this section $(A, \bar A)$ are $sl(2,\mathds{R})$ connections as defined in \cite{Ammon:2013hba}. In particular $\bar A$ in this section has nothing to do with the tensor $\bar A$ defined in \eqref{eq:GA}.} 

It is important that appropriate boundary conditions must be placed on $U$ at the beginning and end of the path. These boundary conditions are chosen such that they are invariant under tangent-space Lorentz rotations, which correspond to the 3-parameter subgroup of gauge transformations in \eqref{Urot} where $L = R^{-1}$. One way to understand why this subgroup is privileged is that it leaves the geometric metric associated to the Chern-Simons gauge fields invariant: the remaining part of $\slt \times \slt$ changes the metric, acting on it (on-shell) as diffeomorphisms. The only $U$ invariant under this privileged subgroup is the identity, and so we impose the boundary conditions $U_i = U_f = \mathds{1}$. 

\subsection{Wilson lines and cosets}
We now want to adapt the discussion above to the case of $\slt \times U(1)$. The $U(1)$ part factors out and will be (easily) dealt with at the end. The non-trivial part then is to find a quantum mechanical system that transforms as a highest-weight representation under only a {\it single} copy of $\slt$. We can then couple this system to a $\slt$ gauge field $B$ and follow the algorithm above to compute the Wilson line. 

Such a system is given by a single particle living on AdS$_2$ rather than AdS$_3$. The isometry group of AdS$_2$ is a single copy of $\slt$. It has been shown that single particle states transform under this $\slt$ as the appropriate highest weight representation \cite{Strominger:1998yg,Spradlin:1999bn}. 

For our purposes, the most efficient way to represent AdS$_2$ is as a coset of $\slt$. We first present a brief review of coset geometry\footnote{Here we immediately specialize to the case of interest, but it should be clear that the discussion applies to any homogenous space.} (see e.g. \cite{Castellani:1999fz}). AdS$_2$ is acted on by $\slt$, and thus one is tempted to pick a reference point $x_0$ in AdS$_2$ and label all other points $x$ by the element of $\slt$ required to move $x_0$ to $x$. This is overcounting, as there is a subgroup $SO(1,1) \subset \slt$ that leaves the reference point $x_0$ fixed, and which should not be used to label points. So we instead understand AdS$_2$ as the coset $\slt / SO(1,1)$. We take the $SO(1,1)$ to be generated by $L_0$, i.e. given any element $U \in \slt$ we may decompose it as
\be
U = g h ~, \label{cosetdecomp}
\ee
where $g = \exp\le(\al L_1 + \beta L_{-1}\ri)$ and $h = \exp(\ga L_0)$. The element $g$ is a coset representative. 

Different $g$'s, modulo the action of $\exp(\ga L_0)$, label different points on a two-dimensional manifold, and there is a canonical way (which we do not review here) to find a $\slt$-invariant metric on this manifold, which is thus seen to be AdS$_2$. The action of $\slt$ is simply via left multiplication on $U$, i.e. $U \to L U$. 

We now generalize the construction above to make the dynamical degree of freedom the coset representative $g$ rather than $U$. This can easily be done by promoting the $L_0$ component of the {\it right} gauge field $\bar A_s$ to a dynamical degree of freedom (which we will now call $\ba_s$) along the worldline. The quantum mechanics along the worldline is now a dynamical gauge theory in its own right. Integration over $\ba_s$ will then gauge away the component of $U$ corresponding to $h$ in the decomposition \eqref{cosetdecomp}, leaving only $g$. This sort of construction is familiar in the context of two-dimensional conformal field theory, although here we are doing it along a one-dimensional worldline. There is still a global symmetry associated with left-multiplication by $\slt$, and as above we couple that global symmetry to an external $\slt$ gauge field which we now call $B$. 

Thus the action is still 
\be
S[U,P,\lam,\ba_s; B] = \int_C ds \le( \Tr\le(P U^{-1} \frac{DU}{ds}\ri) + \lam \le(\Tr(P^2) - c_2\ri)\ri),
\ee
but the covariant derivative is now
\be
\frac{DU}{ds} = \frac{dU}{ds} + B_s U - U \ba_s ~ ,
\ee
where the worldline degree of freedom $\ba_s$ is a number times $L_0$ and the external source $B_s$ is valued in the $sl(2,\mathds{R})$ algebra. 

Finally, we turn now to the choice of boundary conditions on the field $g$. As mentioned above, the key requirement is that the boundary conditions are invariant under a ``privileged'' subgroup of $\slt \times U(1)$, that which leaves the geometric metric \eqref{metident} invariant. This is equivalent to demanding that the subgroup leave invariant the reduced Killing form $\hat{T}_{ab}$. For the choice appropriate to a Euclidean bulk coordinate $X$ as in \eqref{eucBulk}, this is a one-parameter subgroup generated by $L_+ + L_-$, and may be viewed as tangent space $SO(2)$ rotations. This operation acts on $g$ as left-multiplication. As the physical degree of freedom is a coset, ``invariance'' really means that left-multiplication by $L_+ + L_-$ should be equivalent to {\it right}-multiplication by $L_0$, which changes the coset representative but not the coset element itself. Thus our boundary condition $g_{i,f}$ should satisfy $(L_+ + L_-)g_i \propto g_i L_0$, which we can solve to find
\be
g_i = g_f = \exp\le(-i\frac{\pi}{4}(L_+ - L_-)\ri), \label{magicchoice}
\ee
where the solution is ambiguous up to further right-multiplication by $e^{\ga L_0}$. This is the analog of the boundary condition $U_{i,f} = \mathds{1}$ in the $\slt \times \slt$ case. 

\subsection{Equations of motion and on-shell action}
To compute the Wilson line we now need only compute the on-shell action after supplying suitable boundary conditions on $g$ at the two ends of the path. Writing $U = g h$ as in \eqref{cosetdecomp}, we find the equations of motion to be
\begin{align}
h^{-1} g^{-1}\le(\frac{Dg}{ds}\ri)h + \le(h^{-1}\frac{dh}{ds} - \ba_s\ri) + 2 \lam P & = 0 ~, \label{dyneq} \\
\frac{DP}{ds} & = 0 ~,\\
\Tr(P^2) = c_2 ~,\qquad P_{0} \equiv \Tr(P L_0) & = 0 ~, \label{constr}
\end{align}
where the covariant derivatives in question are:
\be
\frac{Dg}{ds} = \frac{dg}{ds} + B_s g ~, \qquad \frac{DP}{ds} = \frac{dP}{ds} + [\ba_s, P] \ .  \label{covdevs}
\ee
The constraints in \eqref{constr} follow from varying with respect to $\lam$ and $\ba_s$ respectively: we stress that integrating out $\ba_s$ means that the $L_0$ component of $P$ must vanish. Note that if we multiply \eqref{dyneq} with $P$ from the left and take the trace, we find that the on-shell action is simply
\be
S_{\rm on-shell}[B] = - 2c_2 \int ds \lam(s) \label{onshellac}
\ee
and we need only determine how $\lam$ varies. 

The main complication in solving these equations arises from the external source $B$. However $B$ will always be flat, so the most efficient way to find a solution is to start in a bulk gauge where $B = 0$ and then perform a gauge transformation on all quantities of interest to arise at the desired solution, as explained in \cite{Ammon:2013hba}. We stress that different choices of $B$ -- even those related by gauge transformations -- are physically inequivalent from the point of view of the worldline, and this procedure is merely a trick to solve the equations of motion. When $B = 0$ these equations are:
\be 
h^{-1} g^{-1}\le(\frac{dg}{ds}\ri)h + \le(h^{-1}\dot{h} - \ba_s\ri) + 2 \lam P = 0  ~, \qquad \frac{dP}{ds} + [\ba_{s}, P] = 0 \ . \label{eom}
\ee
$\ba_s$ is nonzero, as it is still a fluctuating degree of freedom along the worldline, not an external source to be chosen. However we now have the freedom to pick a {\it worldline} gauge for the dynamical gauge field $\ba_s$. We work in a gauge where $\ba_s = 0$, and will show that this is indeed permitted by the boundary conditions of interest. 

In this gauge the most general solution $(U_{\star}(s), P_{\star}(s))$ can be parametrized by a constant element of the group $u_{\star}$ and an element of the algebra $P_{\star}$, and we have
\be
P_{\star}(s) = P_{\star} ~, \qquad U_{\star}(s) = u_{\star} e^{-2\al(s) P_{\star}} ~, \qquad \frac{d\al}{ds} = \lam ~, \qquad \ba_s = 0 \ . 
\ee
Given this reference solution, we now perform a gauge transformation to a flat $B$ field. We denote the resulting solution by $(U(s), P(s))$:
\be
B(x) = L d L^{-1} ~,\qquad U(s) = L(X^{\mu}(s))U_{\star}(s) ~, \qquad P(s) = P_{\star}(s) ~.
\ee
$P$ is not charged under the symmetry associated with left-multiplication (e.g. note that $B$ does not appear in its covariant derivative in \eqref{covdevs}) and so does not change under this gauge transformation. Note that $L(x)$ contains the information of the gravitational background in question. If we now demand that the solution satisfy the boundary conditions at the two ends of the path $U(s_i) = g_i h_i$, $U(s_f) = g_f h_f$, then we obtain eventually the following relation between $P_{\star}$ and the gauge transformation parameter $L(x)$:
\be
e^{-2\Delta\al P_{\star}} = h_f^{-1} g_f^{-1} L(s_f) L(s_i)^{-1} g_i h_i \equiv h_f^{-1} M h_i ~, \qquad \Delta \al = \al(s_f) - \al(s_i) \label{Mdef} ~.
\ee
Now boundary conditions are imposed on the physical degrees of freedom $g_{i,f}$. $h_{i,f}$ then are free parameters of the form $e^{\ga_{i,f} L_0}$. Actually the physics depends only on the difference $h_f h_i^{-1}$, which can in principle be found from integrating the $L_0$ component of \eqref{eom}. Rather than finding it in this way, we note that the role of $h(s)$ is to fluctuate in a manner that allows the $L_0$ component of $P$ to vanish, as is required by the constraint \eqref{constr}. Thus, given $L(x)$, we must pick $h_i h_f^{-1}$ to make sure that $P_{\star}$ above has no $L_0$ component. This is the main practical point of the coset construction. 

In the case of $\slt$ this operation can be implemented explicitly in cases of interest. For illustrative purposes, we perform the computation in the explicit case of the vacuum given by \eqref{ads2vac}. In this case the gauge parameter $L(x)$ is
\be
L(\rho, X) = e^{-\rho L_0} e^{-L_1 X} ~.
\ee
Using the boundary conditions \eqref{magicchoice} we can now explicitly compute $M$ in \eqref{Mdef}. Note that any $M$ can be decomposed as
\be
M = \exp\le(\Sig\le(\nu L_1 - \frac{1}{\nu} L_{-1}\ri)\ri)\exp(\ga L_0) \equiv M_0 \exp(\ga L_0) \label{Mparam} ~.
\ee
This decomposition is helpful as we will pick $h_{f}$ to cancel the $e^{\ga L_0}$ factor at the end. $\ga$ can be obtained using the (easily checked) identity
\be
\sinh\le(\frac{\ga}{2}\ri) = \frac{\Tr(M L_0)}{\sqrt{1 - \Tr(M L_1) \Tr(M L_{-1})}} ~,
\ee
where all traces are taken in the fundamental $2 \times 2$ matrix representation; see (\ref{eq:convsl2}). With the help of this identity and some algebra we can check that $M$ takes the form \eqref{Mparam} where the parameters satisfy
\be
\sinh \Sig = \frac{e^{\rho} \Delta X}{2} ~,\qquad \nu = \sqrt{\frac{2 - i e^{\rho} \Delta X}{2+ i e^{\rho} \Delta X}} ~, \qquad \sinh\le(\frac{\ga}{2}\ri) = \frac{i e^{\rho} \Delta X}{\sqrt{4 + e^{2\rho} (\Delta X)^2}} ~.\label{soln}
\ee
We now pick $h_i = \mathds{1}$ and $h_f = e^{-\ga L_0}$. This is required to guarantee that $P_{\star}$ in \eqref{Mdef} has no $L_0$ component. \eqref{Mdef} becomes
\be
e^{-2\Delta\al P_{\star}} = \exp\le(\Sig\le(\nu L_1 - \frac{1}{\nu} L_{-1}\ri)\ri) ~,
\ee
with $\Sig$, $\nu$ satisfying \eqref{soln}. Taking the trace of both sides we find
\be
2 \cosh(\Delta \al \sqrt{2 c_2}) = \sqrt{4 + (e^{2\rho} \Delta X)^2} ~.
\ee
Finally we take the $e^{\rho}\Delta X \to \infty$ limit -- which means that the interval is very long in units of the cutoff $e^{\rho}$, use the standard Casimir relation $c_2 = 2\Delta_\Phi(\Delta_\Phi - 1)\sim 2\Delta_\Phi^2$, and plug the resulting expression for $\Delta \al$ into \eqref{onshellac} to conclude that
\be
S_{\rm on-shell} = 2\Delta_\Phi \log(e^{\rho} \Delta X),
\ee
where we have taken the large $\Delta_\Phi$ limit. This may seem like a great deal of work to obtain a very simple answer. The essential reason for this is that the Chern-Simons description, while minimal, greatly obscures the geometric description. 

Finally, we return to the $U(1)$ portion. This is trivial: an irreducible unitary representation of a $U(1)$ symmetry is one-dimensional, transforming by multiplication by a phase, and thus there is no need to construct an auxiliary quantum-mechanical system to generate it. If we call the $U(1)$ charge $h$, then the contribution of the $U(1)$ gauge field $\bB$ is simply its integral along the worldline
\be
S_{U(1)} = ih\int_C \bB ~.
\ee
If we plug in the background value \eqref{ads2vac} then we find for the full correlation function
\be
e^{-S_{\rm tot}} \sim \frac{1}{(\Delta X)^{2\Delta_\Phi}} e^{ih\le(\Delta T - \frac{\bL}{L} \Delta X\ri)} ~.
\ee
This is the desired result and it agrees both with (\ref{eq:coorphi}) and (\ref{2p}).
\section{Discussion} \label{sec:conc}

It is undeniable that the power of Conformal Field Theories in two dimensions has provided many insights on the nature of non-perturbative quantum field theory. The exact calculation of entanglement entropy at finite volume (or finite temperature) is one of the miracles possible in this case that has furthered our understanding considerably. Through holography, this result has sparked brand new ways of thinking about quantum gravity. A deep understanding of the meaning and behavior of entanglement entropy in different phases of quantum matter has changed radically the way we think about the entropy of black holes \cite{Ryu:2006bv,Headrick:2010zt} and the nature of space-time itself \cite{Maldacena:2013xja}.

In this work we have shown how to extend these successes both from a standard field theory and a holographic perspective to the realm of Warped Conformal Field Theory. Such powerful results are scarce when it comes to non-relativistic field theories. This makes manifest the importance of WCFTs in possible applications to physical systems. Possible connections with quantum hall physics have been suggested in a related context in \cite{Ryu:2012he} from a CFT perspective and in \cite{Son:2013rqa} from a background geometry perspective. It is a promising open direction to explore this application further.

One particular feature of WCFTs that was of importance in obtaining these results, and in coupling the theory to background geometry, is the existence of two preferred axes. The classical symmetry (\ref{eq:boost}) makes it manifest that the $t$ axis is preferred. But less manifest is the existence of a second preferred direction. The full quantum algebra (\ref{eq:vircov2}) contains anomalies both for the Virasoro and $U(1)$ commutators. A fully covariant writing of this algebra shows that the $U(1)$ anomaly selects another preferred direction in the theory, thus breaking the generalized boost symmetry. This result provides physical motivation for the inclusion of a scaling structure (\ref{eq:scalestru}) in \cite{Hofman:2014loa}.

The main result of this work was, of course, an exact formula for the entanglement entropy of one segment at finite volume in a WCFT, (\ref{eq:oneint}), which we quote again below:
\be\label{eq:oneint2}
S_{\rm EE} = i P_0^{\rm vac} \ell \left( \frac{\bar L}{L} - \frac{ \bar \ell}{\ell}\right) -4 L_0^{\rm vac} \log \left( \frac{L}{\pi \epsilon} \sin \frac{\pi \ell}{L}\right)~.
\ee
The second term is quite reasonable and it agrees with the expected result for a chiral CFT. The first term is, however, more exotic. The fact that it is multiplied by an overall imaginary factor might be upsetting. In any controlled unitary example of WCFT, like the one discussed at length in \cite{Castro:2015uaa}, this term vanishes. In holography, however, this term is generically non zero \cite{Detournay:2012pc}. The entanglement entropy would still be real even in that case as holographic setups predict an imaginary value for $P_0^{\rm vac}$. Notice also that this term is not UV divergent like the standard second term. It is proportional to the volume of the interval and it is only present when there is a misalignment between the segment of interest and the space circle identification ($\frac{\bar L}{L} \neq \frac{ \bar \ell}{\ell}$). This is quite interesting, as we typically see volume terms in the entanglement entropy for mixed states. This result is, however, universal and present for the vacuum (pure) state.

A deeper interpretation of the first term in \eqref{eq:oneint2} is at present lacking. Since it contributes to black hole entropy in holography \cite{Detournay:2012pc}, it might have an important role in the statistical interpretation of black hole thermodynamics. A short discussion on the microcanonical interpretation of these $U(1)$ contributions is presented in \cite{Castro:2015uaa}, but further work in this direction is definitely needed. An important clue that we present in this current work is that this term is actually independent of Renyi replica index (\ref{renyians}), quite differently from the usual behavior in CFT. It would be interesting to understand what is the origin and implications of this behavior. 

One important point is that, as opposed to the case in CFT technology, WCFTs give a geometric meaning to $U(1)$ chemical potentials by providing a torus partition function interpretation. In a WCFT the entanglement entropy of a tilted segment in the cylinder on a pure state maps to a thermal density matrix with potentials turned on both for the Virasoro and the $U(1)$ Kac-Moody algebra. 
This gives a Hilbert space definition of the $U(1)$ charged entanglement entropy discussed, for example, in \cite{Belin:2013uta,Belin:2014mva} for CFTs.

In parallel to the field theory computations discussed above, the same results where obtained from a holographic perspective. This is a necessary check for the Lower Spin Gravity / WCFT correspondence put forward in \cite{Hofman:2014loa}.  Using the the twist field approach in section \ref{sec:twist}, it was easy to reduce the calculation to that of a 2 point function given holographically by the action of a semi-classical particle moving in the warped geometry. As expected, there is a geodesic equation obeyed by these particles and the calculation of the particle action over this preferred path yields the correct result. It is important to stress that this result differs from the expected result in usual Warped AdS Einstein-Chern-Simons gravity, where divergences have been found due to different metric component fall offs \cite{Anninos:2013nja}. Lower Spin Gravity evades these divergences. The reason is that the symmetries of WCFT allow for different couplings of a particle to geometry. While one could attempt an Einstein Gravity holographic description of WCFTs, this assumption is not minimal and implies a different UV behavior responsible for the usual divergences. This problem is also well known in Lifshitz holography where holographic renormalization has proven difficult, see for example \cite{Ross:2011gu}.

It is important to point out that this twist field approach is hard to extend to higher dimensions, where we expect that the holographic calculation is performed by calculating some form of minimal surface as in the Ryu-Takayanagi formula \cite{Ryu:2006bv}. This generalization is at this point not available in the WCFT setup and could be subject to future research.

Lastly, we have also matched this result from a bulk Chern-Simons description. This is the natural language to describe Lower Spin Gravity in the bulk, as advocated in \cite{Hofman:2014loa}. The technology needed to perform this calculation is, in the end, a generalization to coset manifolds of the techniques developed in \cite{Ammon:2013hba} for the $SL(2, \mathbb{R})$ group manifold (AdS$_3$). The full understanding of this setup is of crucial importance as it provides the most natural arena to extend these ideas and provide a fully covariant, democratic and geometric description of higher spin theories and their $W_N$ dual CFTs.

\section*{Acknowledgements}
We would like to thank Dionysios Anninos, G\'{a}bor S\'{a}rosi and Edgar Shaghoulian for discussions related to this project.  We thank G. Stettinger for bringing to our attention an error in an earlier version of this paper. A.C. is supported by Nederlandse Organisatie voor Wetenschappelijk Onderzoek (NWO) via a Vidi grant. This work is part of the Delta ITP consortium, a program of the NWO that is funded by the Dutch Ministry of Education, Culture and Science (OCW).


\appendix

\section{Entanglement entropy in CFT$_2$}\label{app:cft}

In this appendix we review the Rindler method for evaluating entanglement entropy applied to two dimensional CFTs. This is a summary of the general results in \cite{Casini:2011kv} for any dimension applied to two dimensions; see also \cite{Holzhey:1994we,Calabrese:2004eu}. 

\begin{figure}
\includegraphics[width=0.5\textwidth]{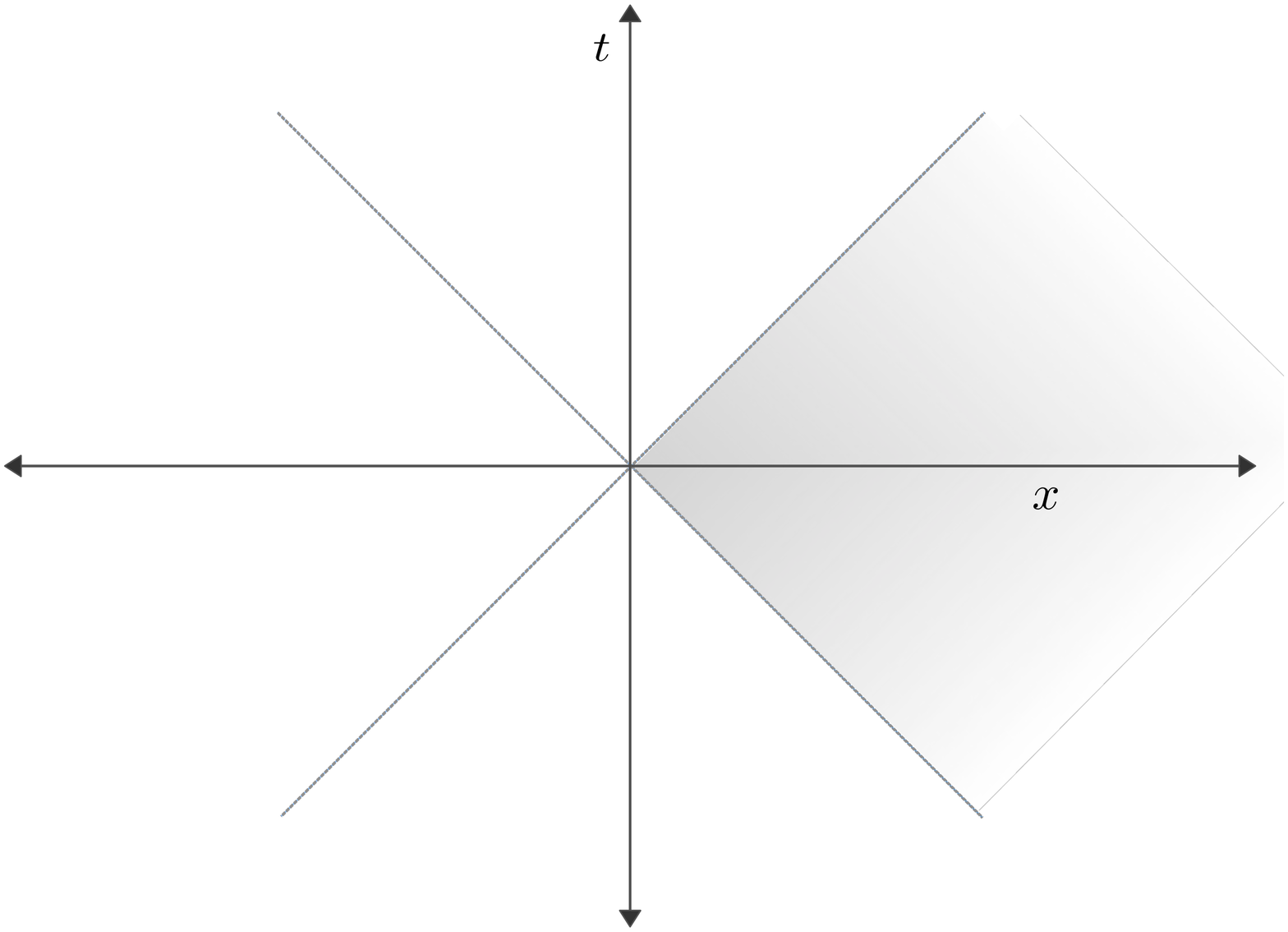}
\includegraphics[width=0.5\textwidth]{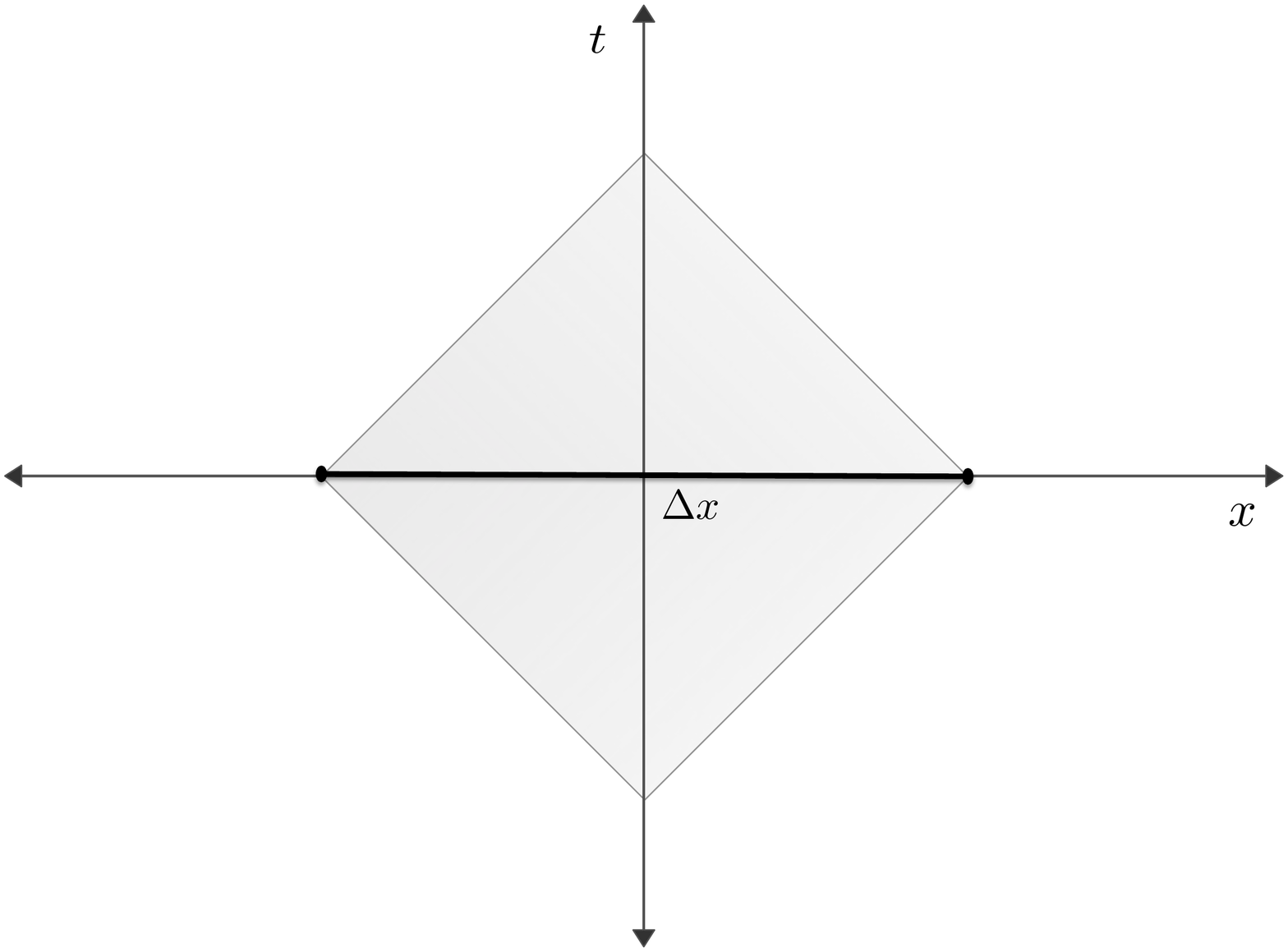}
\caption{{\it Left:} Diagram for entanglement entropy of the half line in a CFT$_2$. Shaded region is the right Rindler wedge; this is the region covered by the coordinates  in \eqref{eq:ypm}. {\it Right:} Diagram for entanglement entropy of a finite segment $\Delta x$ in a CFT$_2$. Shaded diamond is covered by coordinates in \eqref{eq:ypm1}.}\label{fig:cft}
\end{figure}

We define $x^{\pm} \equiv t \pm x$ on the Lorentzian plane, and consider a CFT$_2$ quantized on constant time $t$ slices. To start we will compute the entanglement entropy of the half line ($x>0$) on the vacuum state. Due to Lorentz invariance, this computation can be understood  as the entanglement entropy of the right Rindler wedge. The right Rindler wedge is the intersection of the region $x^+ > 0$ and $x^- < 0$; see figure \ref{fig:cft}.  Coordinates that only cover this patch are 
\be\label{eq:ypm}
x^+ = e^{y^+}~, \qquad x^- = - e^{-y^-}~.
\ee
However, with respect to these coordinates the state of the system looks thermal. This is rather explicit if we write $y^{\pm} \equiv \tau \pm y$ and note that $\tau$ has a natural Euclidean periodicity
\be
\tau \sim \tau + 2\pi i ~. 
\ee
Thus the system in the $(\tau, y)$ coordinates is at finite temperature $T=\beta^{-1} = \frac{1}{2\pi}$.

The density matrix $\rho_{{\rm half}, y^\pm}$ describing the system in the $y^\pm$ coordinates is thermal:
\be
\rho_{{\rm half}, y^\pm} = e^{-\beta H}~, \qquad Z(\beta) \equiv \tr (\rho_{{\rm half},y^\pm})~.
\ee
The basic observation is that this density matrix is related to the original reduced density matrix by the unitary operator $U$ that implements on the Hilbert space the coordinate transformation \eqref{eq:ypm}, i.e.
\be\label{eq:map}
\rho_{{\rm half},x^\pm} = U \rho_{{\rm half},y^\pm} U^{-1}~.
\ee
Since the von Newman entropy  is invariant under unitary transformations,  the entanglement entropy on the half line equals the thermal entropy of the system described by $\rho_{{\rm half},y^\pm}$. Thus the Renyi entropy is simply given by
\be
S_q = \frac{1}{1-q} \log \frac{Z(q\beta)}{Z(\beta)^q}~, \label{renyiform}
\ee
where the denominator arise from the fact that the original density matrix was not normalized. With this equality we find that
\be\label{eq:eehalf}
S_{\rm EE, half}= {\pi c\over 3} L T~,
\ee
 with $T = \frac{1}{2\pi}$ and $L$ is the size of the spatial slice for the Rindler observer in the $y^\pm$ coordinates. Mapping  $L$ to the $x^\pm$ coordinates brings us to an important feature of this method to evaluate entanglement. For the equality between entanglement and thermal entropy to hold, we need to be rather careful with divergent pieces of each observables. Entanglement entropy is divergent at the boundary of the interval and so one needs to introduce a short distance cutoff, whereas thermal entropy is IR divergent due the infinite size of spatial slices for the Rindler observer. Using the conformal mapping \eqref{eq:ypm} we can relate these cutoff procedures: if we place the endpoint at $x_i=\epsilon\to 0$  and $x_f=x_{\rm max}\gg1$, then the domain of $y$ is given by
 \be
 L \equiv \Delta y = \log{x_{\rm max}\over \epsilon}~,
 \ee
and hence
\be
S_{\rm EE, half}= { c\over 6} \log{x_{\rm max}\over \epsilon}~.
\ee

As a second example, we compute the entanglement entropy for a segment of length $\Delta x=R$. As before, due to Lorentz invariance,  this is the entanglement of the 
 \be\label{eq:ypm1}
 x^+= {R\over 2}{e^{y^+}-1\over e^{y^+}+1}~,\quad x^-= {R\over 2}{ e^{y^-}-1\over e^{y^-}+1}~,
 \ee
 where again $y^{\pm} \equiv \tau \pm y$. This observer again has a natural Euclidean periodicity, and hence its density matrix is thermal with   $T=\beta^{-1} = \frac{1}{2\pi}$. The logic follows as above with the only difference being how the UV cutoff $\epsilon$ is related to the IR divergence $L$. Taking $x_i=-{R\over 2}+\epsilon$ and  $x_f={R\over 2}-\epsilon$, from \eqref{eq:ypm1} we find
 \be
 L= 2\log\le({R-\epsilon\over\epsilon}\ri)\sim 2\log\le({R\over\epsilon}\ri)
 \ee
as $\epsilon \to 0$. Using \eqref{eq:eehalf} we find
\be
S_{{\rm EE}}(R)= { c\over 3} \log{R\over \epsilon}~.
\ee


\bibliographystyle{utphys}
\bibliography{wcft}

\providecommand{\href}[2]{#2}\begingroup\raggedright\begin{thebibliography}{10}

\bibitem{Ryu:2006bv}
S.~Ryu and T.~Takayanagi, ``{Holographic derivation of entanglement entropy
  from AdS/CFT},'' \href{http://dx.doi.org/10.1103/PhysRevLett.96.181602}{{\em
  Phys.Rev.Lett.} {\bfseries 96} (2006) 181602},
\href{http://arxiv.org/abs/hep-th/0603001}{{\ttfamily arXiv:hep-th/0603001
  [hep-th]}}.

\bibitem{Ryu:2006ef}
S.~Ryu and T.~Takayanagi, ``{Aspects of Holographic Entanglement Entropy},''
  \href{http://dx.doi.org/10.1088/1126-6708/2006/08/045}{{\em JHEP} {\bfseries
  0608} (2006) 045},
\href{http://arxiv.org/abs/hep-th/0605073}{{\ttfamily arXiv:hep-th/0605073
  [hep-th]}}.

\bibitem{VanRaamsdonk:2010pw}
M.~Van~Raamsdonk, ``{Building up spacetime with quantum entanglement},''
  \href{http://dx.doi.org/10.1007/s10714-010-1034-0,
  10.1142/S0218271810018529}{{\em Gen. Rel. Grav.} {\bfseries 42} (2010)
  2323--2329}, \href{http://arxiv.org/abs/1005.3035}{{\ttfamily arXiv:1005.3035
  [hep-th]}}.
[Int. J. Mod. Phys.D19,2429(2010)].

\bibitem{Swingle:2009bg}
B.~Swingle, ``{Entanglement Renormalization and Holography},''
  \href{http://dx.doi.org/10.1103/PhysRevD.86.065007}{{\em Phys. Rev.}
  {\bfseries D86} (2012) 065007},
\href{http://arxiv.org/abs/0905.1317}{{\ttfamily arXiv:0905.1317
  [cond-mat.str-el]}}.

\bibitem{Maldacena:2013xja}
J.~Maldacena and L.~Susskind, ``{Cool horizons for entangled black holes},''
  \href{http://dx.doi.org/10.1002/prop.201300020}{{\em Fortsch. Phys.}
  {\bfseries 61} (2013) 781--811},
\href{http://arxiv.org/abs/1306.0533}{{\ttfamily arXiv:1306.0533 [hep-th]}}.

\bibitem{Holzhey:1994we}
C.~Holzhey, F.~Larsen, and F.~Wilczek, ``{Geometric and renormalized entropy in
  conformal field theory},''
  \href{http://dx.doi.org/10.1016/0550-3213(94)90402-2}{{\em Nucl.Phys.}
  {\bfseries B424} (1994) 443--467},
\href{http://arxiv.org/abs/hep-th/9403108}{{\ttfamily arXiv:hep-th/9403108
  [hep-th]}}.

\bibitem{Calabrese:2004eu}
P.~Calabrese and J.~L. Cardy, ``{Entanglement entropy and quantum field
  theory},'' \href{http://dx.doi.org/10.1088/1742-5468/2004/06/P06002}{{\em
  J.Stat.Mech.} {\bfseries 0406} (2004) P06002},
\href{http://arxiv.org/abs/hep-th/0405152}{{\ttfamily arXiv:hep-th/0405152
  [hep-th]}}.

\bibitem{Calabrese:2009qy}
P.~Calabrese and J.~Cardy, ``{Entanglement entropy and conformal field
  theory},'' \href{http://dx.doi.org/10.1088/1751-8113/42/50/504005}{{\em
  J.Phys.} {\bfseries A42} (2009) 504005},
\href{http://arxiv.org/abs/0905.4013}{{\ttfamily arXiv:0905.4013
  [cond-mat.stat-mech]}}.

\bibitem{Hofman:2011zj}
D.~M. Hofman and A.~Strominger, ``{Chiral Scale and Conformal Invariance in 2D
  Quantum Field Theory},''
  \href{http://dx.doi.org/10.1103/PhysRevLett.107.161601}{{\em Phys.Rev.Lett.}
  {\bfseries 107} (2011) 161601},
\href{http://arxiv.org/abs/1107.2917}{{\ttfamily arXiv:1107.2917 [hep-th]}}.

\bibitem{Detournay:2012pc}
S.~Detournay, T.~Hartman, and D.~M. Hofman, ``{Warped Conformal Field
  Theory},'' \href{http://dx.doi.org/10.1103/PhysRevD.86.124018}{{\em
  Phys.Rev.} {\bfseries D86} (2012) 124018},
\href{http://arxiv.org/abs/1210.0539}{{\ttfamily arXiv:1210.0539 [hep-th]}}.

\bibitem{Bagchi:2014iea}
A.~Bagchi, R.~Basu, D.~Grumiller, and M.~Riegler, ``{Entanglement entropy in
  Galilean conformal field theories and flat holography},''
  \href{http://dx.doi.org/10.1103/PhysRevLett.114.111602}{{\em Phys. Rev.
  Lett.} {\bfseries 114} no.~11, (2015) 111602},
\href{http://arxiv.org/abs/1410.4089}{{\ttfamily arXiv:1410.4089 [hep-th]}}.

\bibitem{Hosseini:2015gua}
S.~M. Hosseini and A.~Veliz-Osorio, ``{Entanglement and mutual information in
  2d nonrelativistic field theories},''
\href{http://arxiv.org/abs/1510.03876}{{\ttfamily arXiv:1510.03876 [hep-th]}}.

\bibitem{Hosseini:2015uba}
S.~M. Hosseini and Á.~Véliz-Osorio, ``{Gravitational anomalies, entanglement
  entropy, and flat-space holography},''
  \href{http://dx.doi.org/10.1103/PhysRevD.93.046005}{{\em Phys. Rev.}
  {\bfseries D93} no.~4, (2016) 046005},
\href{http://arxiv.org/abs/1507.06625}{{\ttfamily arXiv:1507.06625 [hep-th]}}.

\bibitem{Ryu:2012he}
S.~Ryu and S.-C. Zhang, ``{Interacting topological phases and modular
  invariance},'' \href{http://dx.doi.org/10.1103/PhysRevB.85.245132}{{\em Phys.
  Rev.} {\bfseries B85} (2012) 245132},
\href{http://arxiv.org/abs/1202.4484}{{\ttfamily arXiv:1202.4484
  [cond-mat.str-el]}}.

\bibitem{Son:2013rqa}
D.~T. Son, ``{Newton-Cartan Geometry and the Quantum Hall Effect},''
\href{http://arxiv.org/abs/1306.0638}{{\ttfamily arXiv:1306.0638
  [cond-mat.mes-hall]}}.

\bibitem{Hofman:2014loa}
D.~M. Hofman and B.~Rollier, ``{Warped Conformal Field Theory as Lower Spin
  Gravity},'' \href{http://dx.doi.org/10.1016/j.nuclphysb.2015.05.011}{{\em
  Nucl.Phys.} {\bfseries B897} (2015) 1--38},
\href{http://arxiv.org/abs/1411.0672}{{\ttfamily arXiv:1411.0672 [hep-th]}}.

\bibitem{Anninos:2008fx}
D.~Anninos, W.~Li, M.~Padi, W.~Song, and A.~Strominger, ``{Warped AdS$_3$ Black
  Holes},'' \href{http://dx.doi.org/10.1088/1126-6708/2009/03/130}{{\em JHEP}
  {\bfseries 03} (2009) 130},
\href{http://arxiv.org/abs/0807.3040}{{\ttfamily arXiv:0807.3040 [hep-th]}}.

\bibitem{Anninos:2008qb}
D.~Anninos, ``{Hopfing and Puffing Warped Anti-de Sitter Space},''
  \href{http://dx.doi.org/10.1088/1126-6708/2009/09/075}{{\em JHEP} {\bfseries
  09} (2009) 075},
\href{http://arxiv.org/abs/0809.2433}{{\ttfamily arXiv:0809.2433 [hep-th]}}.

\bibitem{Guica:2011ia}
M.~Guica, ``{A Fefferman-Graham-Like Expansion for Null Warped AdS(3)},''
\href{http://arxiv.org/abs/1111.6978}{{\ttfamily arXiv:1111.6978 [hep-th]}}.

\bibitem{Compere:2013bya}
G.~Compère, W.~Song, and A.~Strominger, ``{New Boundary Conditions for AdS3},''
  \href{http://dx.doi.org/10.1007/JHEP05(2013)152}{{\em JHEP} {\bfseries 05}
  (2013) 152},
\href{http://arxiv.org/abs/1303.2662}{{\ttfamily arXiv:1303.2662 [hep-th]}}.

\bibitem{Ammon:2013hba}
M.~Ammon, A.~Castro, and N.~Iqbal, ``{Wilson Lines and Entanglement Entropy in
  Higher Spin Gravity},'' \href{http://dx.doi.org/10.1007/JHEP10(2013)110}{{\em
  JHEP} {\bfseries 1310} (2013) 110},
\href{http://arxiv.org/abs/1306.4338}{{\ttfamily arXiv:1306.4338 [hep-th]}}.

\bibitem{Castro:2015uaa}
A.~Castro, D.~Hofman, and G.~S\'arosi, ``{Warped Weyl fermion partition
  functions},''
\href{http://arxiv.org/abs/1508.06302}{{\ttfamily arXiv:1508.06302 [hep-th]}}.

\bibitem{Andringa:2010it}
R.~Andringa, E.~Bergshoeff, S.~Panda, and M.~de~Roo, ``{Newtonian Gravity and
  the Bargmann Algebra},''
  \href{http://dx.doi.org/10.1088/0264-9381/28/10/105011}{{\em Class. Quant.
  Grav.} {\bfseries 28} (2011) 105011},
\href{http://arxiv.org/abs/1011.1145}{{\ttfamily arXiv:1011.1145 [hep-th]}}.

\bibitem{Christensen:2013lma}
M.~H. Christensen, J.~Hartong, N.~A. Obers, and B.~Rollier, ``{Torsional
  Newton-Cartan Geometry and Lifshitz Holography},''
  \href{http://dx.doi.org/10.1103/PhysRevD.89.061901}{{\em Phys. Rev.}
  {\bfseries D89} (2014) 061901},
\href{http://arxiv.org/abs/1311.4794}{{\ttfamily arXiv:1311.4794 [hep-th]}}.

\bibitem{Christensen:2013rfa}
M.~H. Christensen, J.~Hartong, N.~A. Obers, and B.~Rollier, ``{Boundary
  Stress-Energy Tensor and Newton-Cartan Geometry in Lifshitz Holography},''
  \href{http://dx.doi.org/10.1007/JHEP01(2014)057}{{\em JHEP} {\bfseries 01}
  (2014) 057},
\href{http://arxiv.org/abs/1311.6471}{{\ttfamily arXiv:1311.6471 [hep-th]}}.

\bibitem{Hartong:2014oma}
J.~Hartong, E.~Kiritsis, and N.~A. Obers, ``{Lifshitz space?times for
  Schrödinger holography},''
  \href{http://dx.doi.org/10.1016/j.physletb.2015.05.010}{{\em Phys. Lett.}
  {\bfseries B746} (2015) 318--324},
\href{http://arxiv.org/abs/1409.1519}{{\ttfamily arXiv:1409.1519 [hep-th]}}.

\bibitem{Hartong:2014pma}
J.~Hartong, E.~Kiritsis, and N.~A. Obers, ``{Schroedinger Invariance from
  Lifshitz Isometries in Holography and Field Theory},''
\href{http://arxiv.org/abs/1409.1522}{{\ttfamily arXiv:1409.1522 [hep-th]}}.

\bibitem{Bergshoeff:2014uea}
E.~A. Bergshoeff, J.~Hartong, and J.~Rosseel, ``{Torsional Newton?Cartan
  geometry and the Schrödinger algebra},''
  \href{http://dx.doi.org/10.1088/0264-9381/32/13/135017}{{\em Class. Quant.
  Grav.} {\bfseries 32} no.~13, (2015) 135017},
\href{http://arxiv.org/abs/1409.5555}{{\ttfamily arXiv:1409.5555 [hep-th]}}.

\bibitem{Geracie:2014nka}
M.~Geracie, D.~T. Son, C.~Wu, and S.-F. Wu, ``{Spacetime Symmetries of the
  Quantum Hall Effect},''
  \href{http://dx.doi.org/10.1103/PhysRevD.91.045030}{{\em Phys. Rev.}
  {\bfseries D91} (2015) 045030},
\href{http://arxiv.org/abs/1407.1252}{{\ttfamily arXiv:1407.1252
  [cond-mat.mes-hall]}}.

\bibitem{Bekaert:2015xua}
X.~Bekaert and K.~Morand, ``{Connections and dynamical trajectories in
  generalised Newton-Cartan gravity II. An ambient perspective},''
\href{http://arxiv.org/abs/1505.03739}{{\ttfamily arXiv:1505.03739 [hep-th]}}.

\bibitem{Hartong:2015xda}
J.~Hartong, ``{Gauging the Carroll Algebra and Ultra-Relativistic Gravity},''
  \href{http://dx.doi.org/10.1007/JHEP08(2015)069}{{\em JHEP} {\bfseries 08}
  (2015) 069},
\href{http://arxiv.org/abs/1505.05011}{{\ttfamily arXiv:1505.05011 [hep-th]}}.

\bibitem{Casini:2011kv}
H.~Casini, M.~Huerta, and R.~C. Myers, ``{Towards a derivation of holographic
  entanglement entropy},''
  \href{http://dx.doi.org/10.1007/JHEP05(2011)036}{{\em JHEP} {\bfseries 05}
  (2011) 036},
\href{http://arxiv.org/abs/1102.0440}{{\ttfamily arXiv:1102.0440 [hep-th]}}.

\bibitem{Dixon:1986qv}
L.~J. Dixon, D.~Friedan, E.~J. Martinec, and S.~H. Shenker, ``{The Conformal
  Field Theory of Orbifolds},''
\href{http://dx.doi.org/10.1016/0550-3213(87)90676-6}{{\em Nucl. Phys.}
  {\bfseries B282} (1987) 13--73}.

\bibitem{Knizhnik:1987xp}
V.~G. Knizhnik, ``{Analytic Fields on Riemann Surfaces. 2},''
\href{http://dx.doi.org/10.1007/BF01225373}{{\em Commun. Math. Phys.}
  {\bfseries 112} (1987) 567--590}.

\bibitem{Cardy:2007mb}
J.~Cardy, O.~Castro-Alvaredo, and B.~Doyon, ``{Form factors of branch-point
  twist fields in quantum integrable models and entanglement entropy},'' {\em
  J.Statist.Phys.} {\bfseries 130} (2008) 129--168,
\href{http://arxiv.org/abs/0706.3384}{{\ttfamily arXiv:0706.3384 [hep-th]}}.

\bibitem{Hung:2011nu}
L.-Y. Hung, R.~C. Myers, M.~Smolkin, and A.~Yale, ``{Holographic Calculations
  of Renyi Entropy},'' \href{http://dx.doi.org/10.1007/JHEP12(2011)047}{{\em
  JHEP} {\bfseries 12} (2011) 047},
\href{http://arxiv.org/abs/1110.1084}{{\ttfamily arXiv:1110.1084 [hep-th]}}.

\bibitem{StettingerThesis}
G.~Stettinger, ``{Twisted Warped Entanglement Entropy},''. Masters thesis (in
  progress), Vienna University of Technology.

\bibitem{Seiberg:1990eb}
N.~Seiberg, ``{Notes on quantum Liouville theory and quantum gravity},''
\href{http://dx.doi.org/10.1143/PTPS.102.319}{{\em Prog. Theor. Phys. Suppl.}
  {\bfseries 102} (1990) 319--349}.

\bibitem{Kutasov:1990sv}
D.~Kutasov and N.~Seiberg, ``{Number of degrees of freedom, density of states
  and tachyons in string theory and CFT},''
\href{http://dx.doi.org/10.1016/0550-3213(91)90426-X}{{\em Nucl. Phys.}
  {\bfseries B358} (1991) 600--618}.

\bibitem{Carlip:1998qw}
S.~Carlip, ``{What we don't know about BTZ black hole entropy},''
  \href{http://dx.doi.org/10.1088/0264-9381/15/11/020}{{\em Class.Quant.Grav.}
  {\bfseries 15} (1998) 3609--3625},
\href{http://arxiv.org/abs/hep-th/9806026}{{\ttfamily arXiv:hep-th/9806026
  [hep-th]}}.

\bibitem{Bianchini:2014uta}
D.~Bianchini, O.~A. Castro-Alvaredo, B.~Doyon, E.~Levi, and F.~Ravanini,
  ``{Entanglement Entropy of Non Unitary Conformal Field Theory},''
  \href{http://dx.doi.org/10.1088/1751-8113/48/4/04FT01}{{\em J. Phys.}
  {\bfseries A48} no.~4, (2015) 04FT01},
\href{http://arxiv.org/abs/1405.2804}{{\ttfamily arXiv:1405.2804 [hep-th]}}.

\bibitem{Lewkowycz:2013nqa}
A.~Lewkowycz and J.~Maldacena, ``{Generalized gravitational entropy},''
  \href{http://dx.doi.org/10.1007/JHEP08(2013)090}{{\em JHEP} {\bfseries 1308}
  (2013) 090},
\href{http://arxiv.org/abs/1304.4926}{{\ttfamily arXiv:1304.4926 [hep-th]}}.

\bibitem{Anninos:2013nja}
D.~Anninos, J.~Samani, and E.~Shaghoulian, ``{Warped Entanglement Entropy},''
  \href{http://dx.doi.org/10.1007/JHEP02(2014)118}{{\em JHEP} {\bfseries 02}
  (2014) 118},
\href{http://arxiv.org/abs/1309.2579}{{\ttfamily arXiv:1309.2579 [hep-th]}}.

\bibitem{Faulkner:2009wj}
T.~Faulkner, H.~Liu, J.~McGreevy, and D.~Vegh, ``{Emergent quantum criticality,
  Fermi surfaces, and AdS(2)},''
  \href{http://dx.doi.org/10.1103/PhysRevD.83.125002}{{\em Phys. Rev.}
  {\bfseries D83} (2011) 125002},
\href{http://arxiv.org/abs/0907.2694}{{\ttfamily arXiv:0907.2694 [hep-th]}}.

\bibitem{Dzhordzhadze:1994np}
G.~Dzhordzhadze, L.~O'Raifeartaigh, and I.~Tsutsui, ``{Quantization of a
  relativistic particle on the SL(2,R) manifold based on Hamiltonian
  reduction},'' \href{http://dx.doi.org/10.1016/0370-2693(94)90549-5}{{\em
  Phys. Lett.} {\bfseries B336} (1994) 388--394},
\href{http://arxiv.org/abs/hep-th/9407059}{{\ttfamily arXiv:hep-th/9407059
  [hep-th]}}.

\bibitem{Maldacena:1998bw}
J.~M. Maldacena and A.~Strominger, ``{AdS(3) black holes and a stringy
  exclusion principle},'' {\em JHEP} {\bfseries 12} (1998) 005,
\href{http://arxiv.org/abs/hep-th/9804085}{{\ttfamily arXiv:hep-th/9804085}}.

\bibitem{Strominger:1998yg}
A.~Strominger, ``{AdS(2) quantum gravity and string theory},'' {\em JHEP}
  {\bfseries 01} (1999) 007,
\href{http://arxiv.org/abs/hep-th/9809027}{{\ttfamily hep-th/9809027}}.

\bibitem{Spradlin:1999bn}
M.~Spradlin and A.~Strominger, ``{Vacuum states for AdS(2) black holes},''
  \href{http://dx.doi.org/10.1088/1126-6708/1999/11/021}{{\em JHEP} {\bfseries
  11} (1999) 021},
\href{http://arxiv.org/abs/hep-th/9904143}{{\ttfamily arXiv:hep-th/9904143
  [hep-th]}}.

\bibitem{Castellani:1999fz}
L.~Castellani, ``{On G / H geometry and its use in M theory
  compactifications},'' \href{http://dx.doi.org/10.1006/aphy.2000.6097}{{\em
  Annals Phys.} {\bfseries 287} (2001) 1--13},
\href{http://arxiv.org/abs/hep-th/9912277}{{\ttfamily arXiv:hep-th/9912277
  [hep-th]}}.

\bibitem{Headrick:2010zt}
M.~Headrick, ``{Entanglement Renyi entropies in holographic theories},''
  \href{http://dx.doi.org/10.1103/PhysRevD.82.126010}{{\em Phys.Rev.}
  {\bfseries D82} (2010) 126010},
\href{http://arxiv.org/abs/1006.0047}{{\ttfamily arXiv:1006.0047 [hep-th]}}.

\bibitem{Belin:2013uta}
A.~Belin, L.-Y. Hung, A.~Maloney, S.~Matsuura, R.~C. Myers, and T.~Sierens,
  ``{Holographic Charged Renyi Entropies},''
  \href{http://dx.doi.org/10.1007/JHEP12(2013)059}{{\em JHEP} {\bfseries 12}
  (2013) 059},
\href{http://arxiv.org/abs/1310.4180}{{\ttfamily arXiv:1310.4180 [hep-th]}}.

\bibitem{Belin:2014mva}
A.~Belin, L.-Y. Hung, A.~Maloney, and S.~Matsuura, ``{Charged Renyi entropies
  and holographic superconductors},''
  \href{http://dx.doi.org/10.1007/JHEP01(2015)059}{{\em JHEP} {\bfseries 01}
  (2015) 059},
\href{http://arxiv.org/abs/1407.5630}{{\ttfamily arXiv:1407.5630 [hep-th]}}.

\bibitem{Ross:2011gu}
S.~F. Ross, ``{Holography for asymptotically locally Lifshitz spacetimes},''
  \href{http://dx.doi.org/10.1088/0264-9381/28/21/215019}{{\em Class. Quant.
  Grav.} {\bfseries 28} (2011) 215019},
\href{http://arxiv.org/abs/1107.4451}{{\ttfamily arXiv:1107.4451 [hep-th]}}.

\end{thebibliography}\endgroup

\end{document}